\documentclass[journal=jpcafh]{achemso}

\usepackage{physics}
\usepackage{graphicx}
\usepackage{dcolumn}
\usepackage{bm}

\usepackage[utf8]{inputenc}
\usepackage{mathtools}
\usepackage[T1]{fontenc}
\usepackage{mathptmx}
\usepackage[utf8]{inputenc}
\usepackage{gensymb}
\usepackage{graphicx}
\usepackage{subcaption}
\usepackage{amsmath}
\usepackage{amssymb}
\usepackage{dcolumn}
\usepackage{bm}
\usepackage{xr}
\usepackage{cleveref}
\usepackage{soul}
\usepackage{color}
\usepackage{ulem}
\usepackage[usenames,dvipsnames]{xcolor}
\usepackage{epstopdf}
\AppendGraphicsExtensions{.tif}

\title{Surface-Adsorbed Nanodroplets of Symmetric Diblock Copolymers Form Versatile and Stimuli-Responsive Nanostructures}
\author{Artem Petrov}
 \email{aipetrov@mit.edu}
	\affiliation{Department of Chemical Engineering, Massachusetts Institute of Technology, Cambridge, Massachusetts 02139, United States}
 \author{Guillermo A. Hern\'{a}ndez-Mendoza}
	\affiliation{Department of Materials Science and Engineering, Massachusetts Institute of Technology, Cambridge, Massachusetts 02139, United States}
 \author{Alfredo Alexander-Katz}
\email{aalexand@mit.edu}
	\affiliation{Department of Materials Science and Engineering, Massachusetts Institute of Technology, Cambridge, Massachusetts 02139, United States}
	\date{\today}

\begin{document}

 \begin{figure*}[tb]
    \centering
	\includegraphics[width=0.4\textwidth]{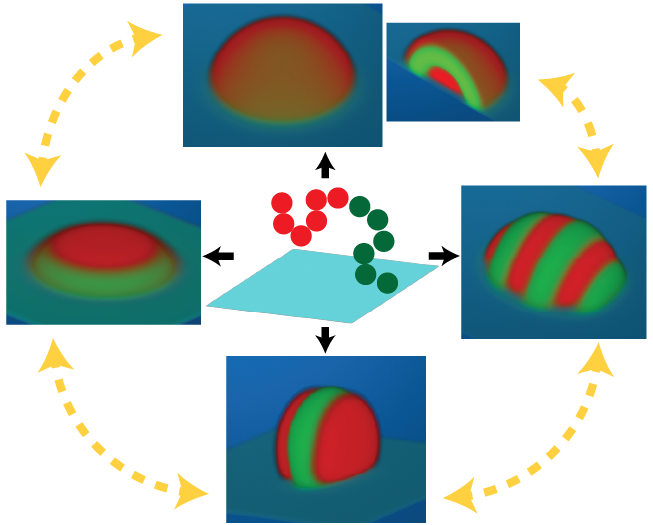} \caption*{}
    \label{}
\end{figure*}
	
\begin{abstract}
Block copolymers often create droplets when placed on a substrate. Such nanostructured droplets can be arranged into regular microstructured arrays, thereby forming hierarchically organized materials that can be used in microelectronics, plasmonics, sensing, photonics, metamaterials production, and even cryptography. However, it is unclear if such materials can be stimuli-responsive, \textit{i.e.}, be able to change their nanostructure on a single droplet level upon applying external stimuli. In this work, we discovered that small (10-100 nm) surface-adsorbed droplets of symmetric diblock copolymers can form a multitude of different externally switchable nanostructures. We obtained a near-equilibrium, comprehensive 4D diagram of droplet morphologies by performing large-scale self-consistent field theory (SCFT) calculations under various wetting and phase separation conditions. The SCFT modeling was augmented with a computational algorithm that established an equilibrium droplet morphology in a given system without assuming potentially equilibrium structures prior to simulation. The discovered droplet nanostructures agreed excellently with previously published experimental data. Crucially, we showed that direct and reversible transitions between different droplet morphologies are possible upon changing the interaction strength between components, which can be tuned externally in experiments by adding surfactants or controlling temperature. We confirmed experimental realizability of such stimuli-responsiveness by modeling surfactant addition that led to a switch between droplet nanostructures. This work demonstrates that even the simplest symmetric diblock copolymers are able to produce versatile and stimuli-responsive structures on a surface when confined to a small nanodroplet. This opens the possibility to produce smart coatings with externally switchable hierarchical micro- and nanostructures.

\end{abstract}

\maketitle


Block copolymers are macromolecules consisting of several different monomer species organized into blocks along the chain. Such molecules can form a vast diversity of nanostructured materials \cite{mai2012self}. This ability of block copolymers to self-organize into versatile and well-defined nanoscale patterns on relatively large scales makes such macromolecules a very promising candidate for a wide variety of nanotechnology applications \cite{sinturel2015high}. One of the most important fields where block copolymers can be potentially widely implemented is the microelectronics industry, since the self-assembly of block copolymers into on-demand nanopatterns on a surface can be an alternative to the traditional lithographic techniques, forming smaller and more controlled nanofeatures at a significantly lower cost \cite{huang2021block,sinturel2013solvent,morris2015directed,jeong2013directed}.

In order to form such surface self-assembled nanostructures, the substrates are usually coated with a thin film of block copolymers. The control over the nanoscale morphology of block copolymer thin films was extensively investigated over the past two decades experimentally and computationally \cite{cheng2018templated,ding2019emergent,gadelrab2018limits,huang2019dissipative,morris2015directed,hu2014directed,jeong2013directed,mickiewicz2010enhancing}. Besides the microelectronics applications, such films can be used as surfaces with advanced adhesive, hydrophobic or antireflective properties \cite{huang2021block,sinturel2013solvent}. However, one of the biggest factors limiting the creation of such nanostructured coatings remains the tendency of block copolymers to wet the substrate weakly enough under certain conditions (for example, after thermal or solvent annealing), which results in film dewetting and the formation of micro- and nano-sized droplets of block copolymers \cite{sinturel2013solvent,brassat2020nanoscale,ferrarese2018hierarchical,ferrarese2020tailored,green2001block,choi2012dewetting,farrell2011surface,kim2008hierarchical}. Interestingly, instead of preventing this process, researchers suggested to use it in order to create a novel class of hierarchically organized materials. Such systems are obtained by annealing block copolymer thin films on pre-patterned substrates \cite{kim2008hierarchical,farrell2011surface,brassat2020understanding,kim2023dewetting,kim2010micropatterns}. This process leads to the formation of a regular microstructured array of block copolymer droplets. Droplets themselves, in turn, undergo disorder-to-order transition and form their own nanoscale structure, which endows the entire coating with the hierarchical organization. Importantly, such droplet-based coatings can be used in the microelectronics applications \cite{farrell2011surface} as well as in plasmonics, photonics, sensing, metamaterials production, and even cryptography \cite{ferrarese2020tailored,ferrarese2018hierarchical,torun2021physically,murataj2021hyperbolic}.

It is worth mentioning that the substrate pre-patterning allows a relatively straightforward programming of droplet microscale arrangement independently of the internal nanoscale structure of the droplets \cite{brassat2020understanding,kim2008hierarchical,ferrarese2018hierarchical}. In other words, it is potentially possible to tune the nanostructure of such surface materials on the level of one droplet orthogonally to the control of the microscopic organization of such coatings. However, this droplet nanostructure tuning remains currently underdeveloped. A handful of works demonstrated the possibility of controlling the nanoscale morphology in related hierarchical surface materials, for instance, in an array of nanoholes \cite{brassat2020understanding,kim2009ordered}. However, for droplet-based coatings, it is an open question whether it is possible to obtain a stimuli-responsive hierarchically structured material. In turn, droplet-based hierarchical coatings that can switch between different nano-level organizations upon changing external conditions can become the next-generation smart materials with predictable and controllable chemical and topographical properties.

The current insufficient ability to form stimuli-responsive hierarchical block copolymer surface coatings stems mostly from the lack of understanding of a single block copolymer droplet organization. Since it is very hard to perform such systematic investigation directly in the microstructured array of droplets, several works studied the nanostructure of a block copolymer droplet as a single entity or in a disordered array of droplets \cite{kim2009droplets,croll2006droplet,croll2009spreading,mcgraw2011dynamics,ilton2014quantized,hur2015interplay,cohen2014interfacial,man2012block,wang2019self,choi2012dewetting}. In solution, the single nanoparticle morphologies were investigated extensively in experiments \cite{brisson2024nanoscale,klinger2014facile,deng2014janus,deng2015soft,jang2013striped,hu2021light} and, recently, in simulations \cite{petrov2025symmetric,he2024phase,huang2024design,chi2011soft,deng2021transformable,yan2016self}. However, the studies on the single surface-adsorbed block copolymer droplets are much more limited.
The vast majority of these works considered large droplets having size ranging from hundreds of nanometers to tens of microns \cite{kim2009droplets,croll2006droplet,croll2009spreading,mcgraw2011dynamics,ilton2014quantized,choi2012dewetting}. These studies revealed that microdroplets of symmetric diblock copolymers, which consist of two monomer species arranged into two blocks of equal volume fraction, had nontrivial terrace-like shape. However, the overall droplet morphology depended weakly on the interaction parameters \cite{kim2009droplets}, which pointed to a potentially poor tunability and stimuli-responsiveness of the nanostructure in such large droplets.

On the other hand, experimental works identified the possible formation of hierarchical surface assemblies with smaller droplets having size ranging between tens to hundreds of nanometers, which had nontrivial nanoring-like structure \cite{farrell2011surface}. \textit{In solution}, small block copolymer nanoparticles having size on the order of tens of nanometers were shown to have a highly versatile and controllable nanostructure \cite{petrov2025symmetric}. However, very few experimental and computational studies investigated the structure of \textit{surface-adsorbed} block copolymer droplets having small ($10-100$ nm) size. Interestingly, they discovered the formation of two other droplet structures: striped perpendicular domains and decorated disks \cite{hur2015interplay,cohen2014interfacial,man2012block}. For more exotic graft polymers, onion-like nanostructure was predicted for small nanodroplets on a surface \cite{wang2019self}. Therefore, these studies pointed out the ability of nanosized droplets of block copolymers to potentially form unusual structures on a surface.
However, these rare works did not investigate the structural versatility of block copolymer nanodroplets, the transitions between them, or their stimuli-responsiveness.
Instead, they mostly focused on the properties of a particular nanostructure and its dependence on droplet size (which is also very hard to tune externally without re-making the material).
Moreover, despite the propensity of block copolymers to form metastable states \cite{tsai2022using,bosse2006defects}, these studies had very limited analysis of equilibration of the discovered structures. Therefore, the phase behavior of single block copolymer nanodroplets remains largely underexplored despite the potential of such objects to be a platform for stimuli-responsive and versatile hierarchical surface coatings.

In this work, we addressed this problem by constructing a comprehensive, four-dimensional morphological diagram of surface-adsorbed nanodroplets of symmetric diblock copolymers. We obtained this diagram by varying available interaction strength parameters in the system; they can be controlled externally in an experiment without re-making the material. The size of the droplets was fixed to tens of nanometers for monomers typically used in experiments. We performed large-scale self-consistent field theory (SCFT) calculations ($\approx 2\times 10^4$ individual system calculations), since SCFT proved to be rather accurate in reproducing the experimentally discovered morphologies of block copolymer droplets in solution \cite{petrov2025symmetric}. After the calculations, we developed and implemented an algorithm that corrected the SCFT calculation output and generated a near-equilibrium morphological diagram of block copolymer nanodroplets without requiring prior knowledge of potentially equilibrium droplet structures. As a result, we discovered that the simplest symmetric diblock copolymers can form a large variety of equilibrium nanostructured droplets, and the transitions between these structures can be realized by tuning the interaction strength between components in the system. This, in turn, can be realized experimentally \textit{via} the addition of surfactants or \textit{via} temperature/pH control. We demonstrated the viability of such stimuli-responsiveness of droplets by simulating the addition of amphiphilic molecules that modeled surfactants and observing the switch between nanodroplet structures. Our results were supported by the previously published experimental data \cite{hur2015interplay}, and the pathways for obtaining newly discovered nanodroplet morphologies in experiment were proposed. As a result, we demonstrated that confinement of the simplest diblock copolymers into $10-100$ nm sized droplets gives rise to their highly versatile and externally controllable phase behavior, which can provide good external tunability of the nanostructure of hierarchical surface coatings.

\section{Results and Discussion}

\begin{figure}[H]
  \centering
  \includegraphics[width=0.65\linewidth]{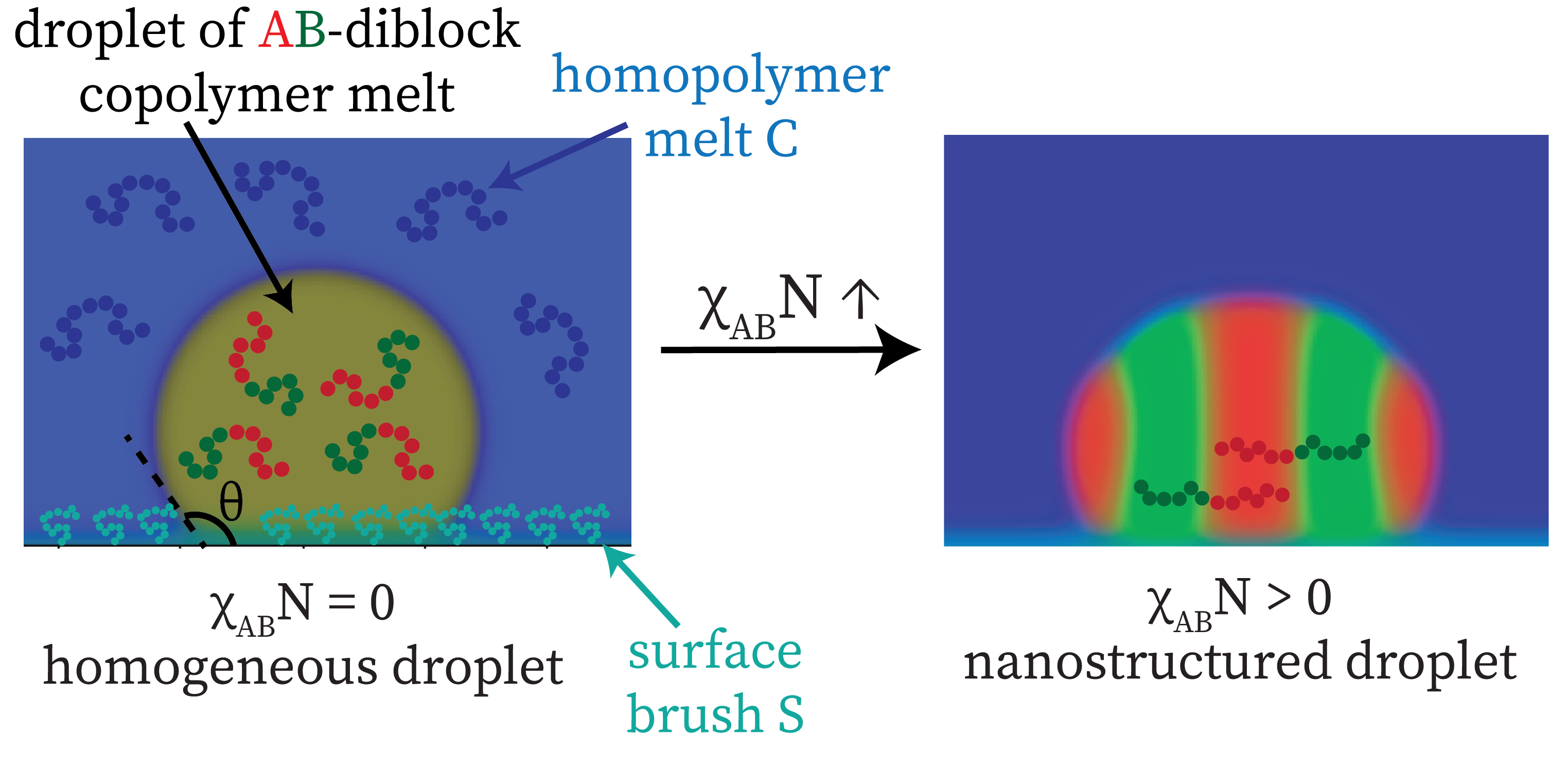}
  \caption{Sketch of the studied system. Symmetric diblock copolymers formed a droplet on the surface; $A$ and $B$ segments are shown in red and green, respectively. Surface was covered by the short brush $S$ colored in light cyan. Melt of homopolymers $C$ that modeled the surrounding medium (vacuum, gas, solvent or solvent vapor) is colored in dark blue. Before SCFT calculations, we prepared a homogeneous droplet ($\chi_{AB}=0$) having a certain contact angle $\theta$ (left). After this, $\chi_{AB}N$ was set to a target value and a nanostructured droplet was formed (an example of such structure is shown on the right).}
  \label{fig:sketch}
\end{figure}

We studied small surface-adsorbed droplets of fixed volume formed by symmetric diblock copolymers. We performed SCFT calculations in two-dimensional (2D) (section 2.1) and 3D (section 2.2) space. The characteristic droplet size was on the order of tens of nanometers for typical block copolymers (BCP) with Kuhn length $\approx 1$ nm and degree of polymerization $\propto 10^2$ (see Methods). The medium surrounding a droplet (gas, vacuum, solvent, or solvent vapor) was modeled by homopolymers $C$ having the same length as BCP chains. The surface (crosslinked random copolymers of various composition in typical experiments \cite{hur2015interplay}) was modeled by tethered short homopolymers $S$ (Figure \ref{fig:sketch}). The following six effective Flory-Huggins parameters defined interactions between system components: $\chi_{AB}N$, $\chi_{AC}N$, $\chi_{BC}N$, $\chi_{AS}N$, $\chi_{BS}N$, and $\chi_{CS}N$. We fixed $\chi_{BC}N=18$ for all studied systems. Parameters $\chi_{AS}N$, $\chi_{BS}N$, and $\chi_{CS}N$ determined the surface tensions between the corresponding species \cite{hur2012chebyshev}. As a result, their combination set the contact angle $\theta$ of a structureless diblock copolymer droplet (\textit{i.e.}, at $\chi_{AB}=0$, see Figure \ref{fig:sketch}) according to equations obtained empirically in Supporting Information (SI) section 1. Moreover, at a fixed $\theta$, one can vary $\chi_{AS}N$ or $\chi_{BS}N$ to tune $Pref_A$: the preference of the surface to $A$ blocks in a structureless droplet (at $\chi_{AB}=0$). We defined $Pref_A$ as $Pref_A=\phi_A^{surf}-f$, where $\phi_A^{surf}$ is the fraction of $A$-type segments in a thin layer near the surface inside a structureless BCP droplet, and $f=0.5$ is the BCP composition (see SI section 1 for details). As a result, the three parameters $\chi_{AS}N$, $\chi_{BS}N$, and $\chi_{CS}N$ were reduced to two effective physically relevant independent parameters $\theta$ and $Pref_A$ characterizing the strength of BCP-surface wetting and the preference of the surface to one of the blocks, respectively.

\subsection{2D Space SCFT Calculations}

First, we studied the nanostructure of droplets in 2D space. We varied $Pref_A$ in the $Pref_A\in[-0.064,0.157]$ range (from weakly $B$-preferential to strongly $A$-preferential substrates). For each value of $Pref_A$, we obtained morphological diagrams of nanodroplets by varying $\chi_{AB}N$, $\chi_{AC}N$, and $\theta$. We varied the interaction parameters $\chi_{AB}N$ and $\chi_{AC}N$ in the following ranges: $\chi_{AB}N\in[10,50]$ (from very weak to very strong $A$-$B$ segregation), $\chi_{AC}N\in[0,\chi_{BC}N=18]$ (from full compatibility of block $A$ with the surrounding medium to block $A$ being as incompatible with the medium as block $B$). As shown in ref. \cite{petrov2025symmetric}, this range was sufficient to obtain all equilibrium experimentally attainable structures for BCP nanoparticles in solution of the studied volume. The value of $\theta$ was set to three values: $\theta\approx67\degree$, $\theta\approx92\degree$, $\theta\approx141\degree$ in order to span the wetting conditions in a wide range.

\subsubsection{Surfaces Without Preference to Any of the Blocks}

\begin{figure}[h!]
\centering
  \begin{subfigure}{0.49\textwidth}
	\includegraphics[width=\linewidth,height=\textheight,keepaspectratio]{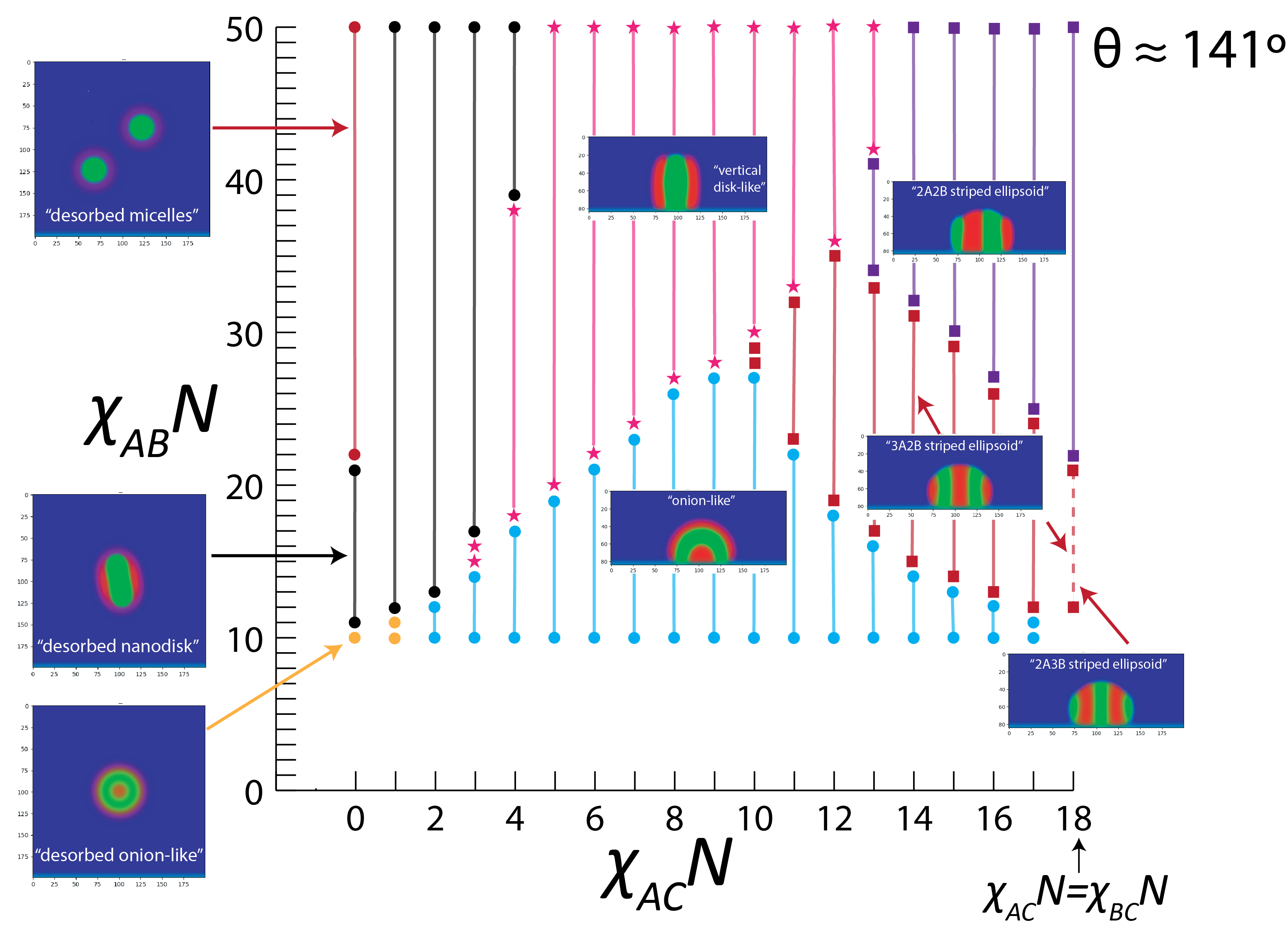}
	\caption{}
	\end{subfigure}
	\begin{subfigure}{0.49\textwidth}
	\includegraphics[width=\linewidth,height=\textheight,keepaspectratio]{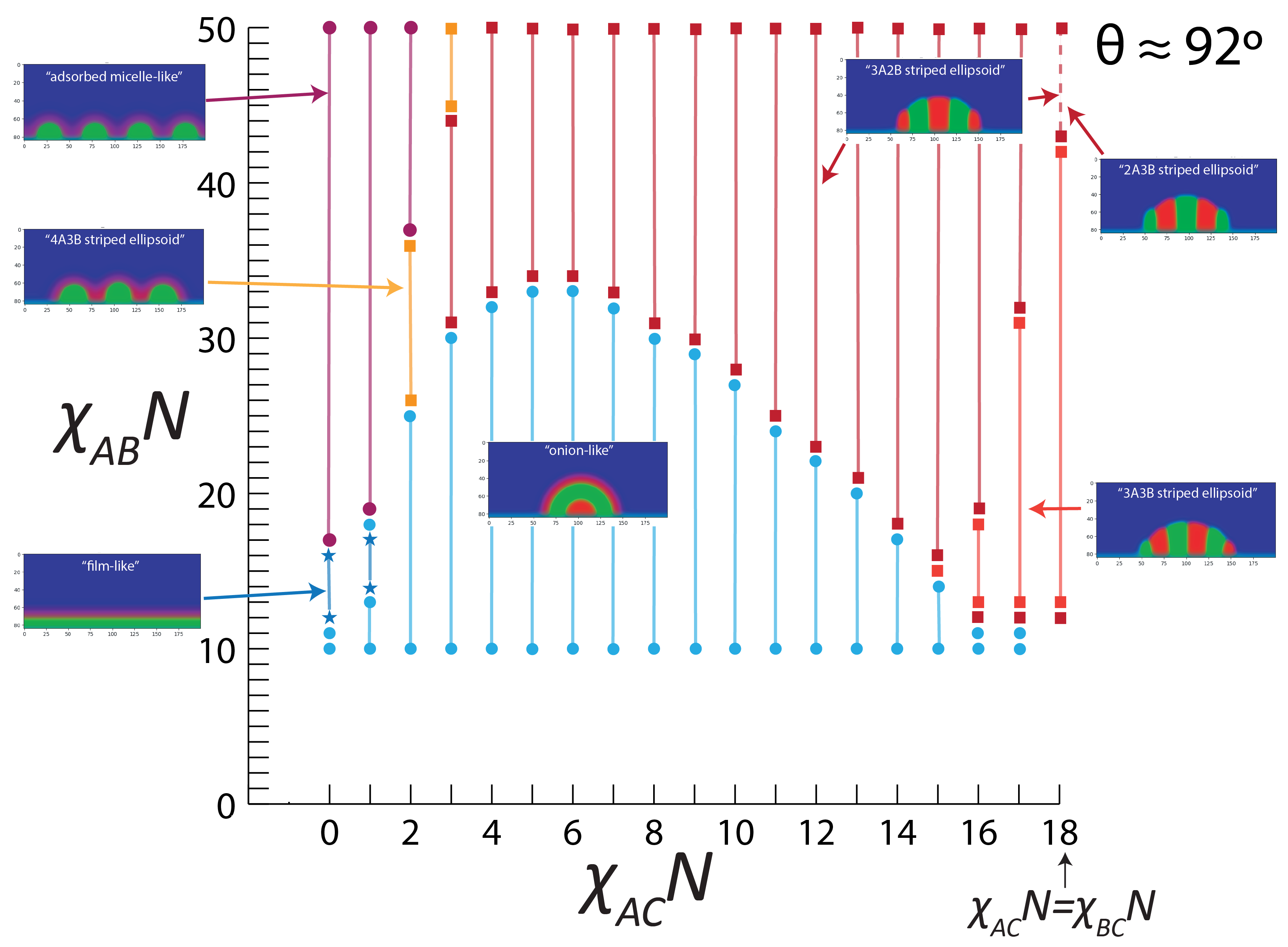}
	\caption{}
	\end{subfigure}
	\begin{subfigure}{0.49\textwidth}
	\includegraphics[width=\linewidth,height=\textheight,keepaspectratio]{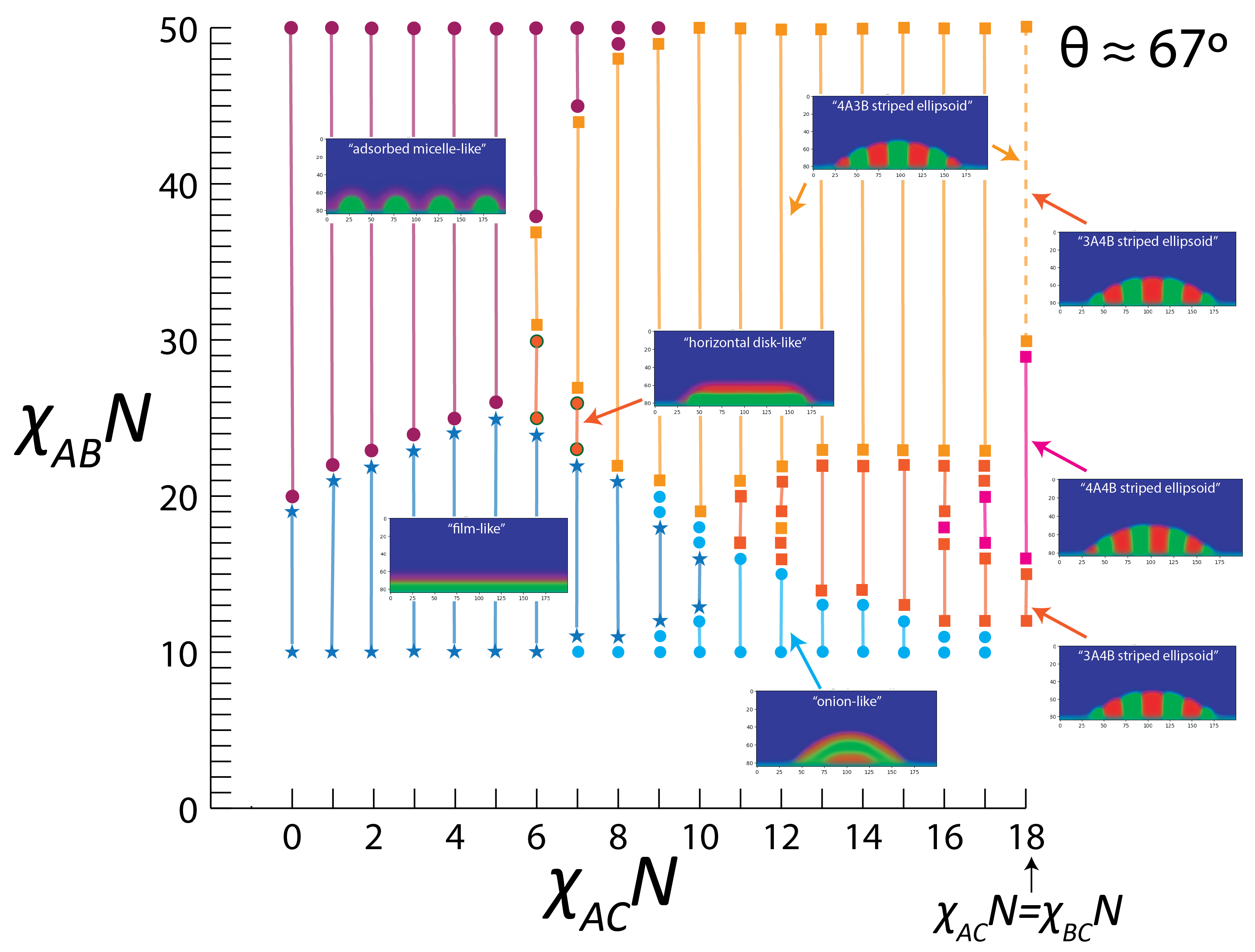}
	\caption{}
	\end{subfigure}
  \caption{Equilibrium morphological diagram of nanodroplets adsorbed at a surface with no preference towards any of the blocks ($Pref_A\approx 0$). SCFT calculations were performed in 2D space. Diagrams (a), (b), and (c) were obtained at $\theta\approx 141\degree$, $\theta\approx 92\degree$, and $\theta\approx 67\degree$, respectively. Points of different colors represent the boundaries of different morphologies at a fixed $\chi_{AC}N$. Continuous lines between the boundary points show that this morphology was obtained at all values of $\chi_{AB}N$ between these points. A characteristic snapshot for each structure as well as its short name are shown for each morphological region. $A$ blocks are shown in red, $B$ blocks are colored green, the surface-tethered $S$ homopolymers are shown in light blue, and homopolymers $C$ modeling the surrounding medium are colored dark blue. Dashed lines represent the alternating presence of two morphologies between the boundary points at the given $\chi_{AC}N$. Since these morphologies did not differ in the number of lamellar layers but only in their order (for example, "$3A2B$" and "$2A3B$" striped ellipsoidal particle in (a)), and since $\chi_{AC}=\chi_{BC}$ along the dashed lines, there was no preference towards any of the two morphologies and they alternated randomly at these conditions.}
\label{fig:1}
\end{figure}

Figure \ref{fig:1} shows the equilibrium morphological diagrams for the symmetric diblock copolymer nanodroplets adsorbed at a nonpreferential surface ($\chi_{AS}=\chi_{BS}$, $Pref_A\approx 0$). These diagrams were obtained by calculating an "initial" diagram using SCFT real-space method \cite{sides2003parallel,drolet1999combinatorial} (Figure S2) and subsequently correcting the diagram according to the iterative correction algorithm (ICA). During ICA, we selected a structure lying near the transition surface between morphological regions and carried out SCFT calculations initialized from nearby structures having different morphology (at the interaction parameters of the selected border structure). After carrying out such calculations for all boundary structures, the borders between morphological regions were updated based on the lowest free energy state (see section 4.2 for more detailed description of the ICA). The "initial" diagram used as the ICA input and the diagram after the first iteration of ICA are shown in Figure S2 and S3, respectively. The comparison of these two Figures shows that ICA changes the morphological diagram dramatically iteration to iteration, shifting morphological boundaries, removing metastable structures, and producing new structures at early iterations. However, the algorithm converged to the equilibrium diagram (Figure \ref{fig:1}) after 37 iterations (7765 individual real-space SCFT calculations). The equilibrium nature of this diagram was confirmed by taking all obtained structures as initial states and carrying out SCFT calculations in the center of each equilibrated morphological region in Figure \ref{fig:1} (see section 4.2 for details). Comparing the "initial" and "final" diagrams (Figure S2 and \ref{fig:1}, respectively), one can observe that the number of structures approximately halved after ICA completion. This, in turn, indicates an abundant formation of metastable states of droplets on non-preferential surfaces after initial real-space SCFT runs. Therefore, is was necessary to perform the ICA in this system to remove the metastable structures in order to correctly study the nanodroplet equilibrium phase behavior.

Figure \ref{fig:1}a shows the equilibrium structures of nanodroplets when BCP wetted the surface weakly (\textit{i.e.}, $\theta\approx141\degree$). When blocks $A$ had a strong preference to the surrounding medium compared to blocks $B$ (\textit{i.e.}, $\chi_{AC}N\lesssim 4$), the droplets preferred not to adsorb at the surface at equilibrium for almost any block-block separation ($\chi_{AB}N$). Instead, they desorbed forming structures as if placed in a solution. The equilibrium structures included spherical micelles as observed experimentally and predicted theoretically for symmetric diblock copolymers in good solvent conditions \cite{lee2020interfacial}. Moreover, a small region of onion-like particles was observed for very weak block-block separation. In addition, nanodisks were ubiquitously formed when solvent was strongly preferential to blocks $A$ and block-block separation was strong enough as observed theoretically \cite{petrov2025symmetric} and experimentally \cite{deng2014janus}.

On the other hand, when the asymmetry of block-medium interactions was not so large ($\chi_{AC}N\gtrsim 4$), the nanodroplets preferred to adsorb at the surface at equilibrium. When $\chi_{AC}N$ was close to $\chi_{BC}N$, the asymmetry of block-medium interaction became weak enough, and the dominant equilibrium structure was the striped ellipsoidal nanodroplet with roughly similar exposure of $A$ and $B$ blocks to the surrounding medium. Such nanodroplets were previously obtained experimentally (see section 2.2 and ref. \cite{hur2015interplay}). Upon an increase of block-block segregation strength $\chi_{AB}N$, the number of lamellar periods in a nanodroplet tended to decrease, which resulted from a larger thickness of the period and an increased energetic penalty for forming the small highly-curved nanodomains at the edges of the droplets. Interestingly, at moderate $\chi_{AB}N\leq35$, the symmetric "$2A2B$" striped ellipsoidal droplets transitioned into asymmetric "$3A2B$" droplets upon decreasing $\chi_{AC}N$, which resulted from an increased asymmetry between the interactions of two blocks with the surrounding medium. At higher $\chi_{AB}N$ or at lower $\chi_{AC}N$, the striped ellipsoidal structures transitioned into a novel, previously unobserved "standing disk" structure, where the chains organized into a bilayer standing perpendicularly to the surface with $A$ blocks preferentially contacting the medium. The formation of this structure was alleviated by (i) a high block-block separation, which penalized curving and promoted the formation of a flat bilayered structure and (ii) optimal block-medium interaction asymmetry, which was strong enough to promote the migration of the $A$ blocks to the periphery of the structure and, at the same time, not large enough to promote desorption.

Finally, at low $\chi_{AB}N$, the formation of curved lamellae was not heavily penalized, which resulted in the formation of onion-like adsorbed structures at almost all $\chi_{AC}N$ values. Such structures also were not observed experimentally before \cite{hur2015interplay}. Setting $\chi_{AC}N$ to an intermediate value promoted the formation of such structures; this happened, on one hand, due to the absence of droplet desorption ($\chi_{AC}N$ was large enough) and, on the other hand, due to a considerable asymmetry of interaction between the blocks and the surrounding medium that attracted the $A$ blocks to the medium ($\chi_{AC}N$ was smaller than $\chi_{BC}N$).

When BCP wetted the surface stronger ($\theta\approx 92\degree$, Figure \ref{fig:1}b), the trends remained qualitatively similar except at none of the studied interaction parameters the formation of desorbed nanodroplets was allowed at equilibrium. Instead, the onion-like and the "$3A2B$" striped ellipsoidal droplets were dominating the morphological diagram. The formation of onion-like nanodroplets was promoted at low $\chi_{AB}N$ and intermediate $\chi_{AC}N$ for the reasons discussed in the previous paragraph. The "$3A3B$" striped ellipsoidal nanodroplet transitioned into the "$3A2B$" striped ellipsoid upon increasing $\chi_{AB}N$, since the number of lamellar layers tended to decrease upon increasing block-block separation (see previous paragraph). Moreover, the formation of the latter type of the droplet was facilitated at lower $\chi_{AC}N$ since blocks $A$ are naturally more exposed to the medium in the "$3A2B$" striped ellipsoidal structure compared to the "$3A3B$" state.

Interestingly, at low $\chi_{AC}N$, the striped ellipsoidal and onion-like droplets transitioned into structures that appeared to fully cover the surface. At low $\chi_{AB}N$, BCP covered the entire surface with a uniform structured film; however, calculations in a box with a larger horizontal ($x$) dimension (Figure S4) showed that this is most likely an artifact stemming from a finite calculation box size in Figure \ref{fig:1}. In boxes with larger $x$-dimension (with the volume of the droplet being fixed), such films formed a droplet of horizontal disk type (also obtained at $\theta\approx67\degree$, see Figure \ref{fig:1}c). At higher $\chi_{AB}N$, BCPs formed an array of surface-adsorbed micelles. Calculations in Figure S4 showed that an increase of the $x$-dimension of a calculation cell led to an increased distance between the micelles, which suggested that symmetric diblock copolymers with such interaction parameters would form an array of isolated adsorbed micelles on an infinite surface at equilibrium. Possibly, the uniform or undulated film-like structures observed in Figure \ref{fig:1} may form on an infinite surface if the \textit{volume fraction} (and not volume) of BCP is kept constant upon an increase of the $x$-dimension of the cell; however, the equilibration of such states may be affected by fluctuation-induced instabilities not taken into account in SCFT.

Finally, at low value of $\theta\approx67\degree$ (Figure \ref{fig:1}c), BCP wetted the surface relatively strongly. As a result, the adsorbed micelle-like and film-like structures appeared in the diagram at low enough $\chi_{AC}N$. At $\chi_{AC}N=\{6,7\}$ and $\chi_{AB}N\in[23,30]$, the formation of horizontally oriented disk-like nanodroplet was observed. Such structures were previously obtained in experiments \cite{hur2015interplay} and result from (i) strong enough wetting of the BCP and the surface, (ii) low enough $\chi_{AC}N$ that promoted the full exposure of $A$ blocks to the surrounding medium, and (iii) low enough $\chi_{AB}N$. At higher $\chi_{AB}N$ and $\chi_{AC}N$, such adsorbed nanodisks transitioned into "$4A3B$" striped ellipsoidal nanodroplets.

As a result, we obtained a comprehensive description of equilibrium phase behavior of small BCP nanodroplets adsorbed on neutral surfaces. By tuning the block-block segregation, asymmetry of interaction between the blocks and the surrounding medium, as well as BCP wetting strength, one can obtain a large variety of nanostructured droplets: striped ellipsoidal structures with tunable number of layers, onion-, micelle-, and disk-like adsorbed droplets that expose only one type of segments to the medium, and the anisotropic bilayered vertically-oriented disk-like structure.

\subsubsection{Surfaces With Non-Zero Preference to $A$ Blocks}

\begin{figure}[h!]
\centering
  \begin{subfigure}{0.49\textwidth}
	\includegraphics[width=\linewidth,height=\textheight,keepaspectratio]{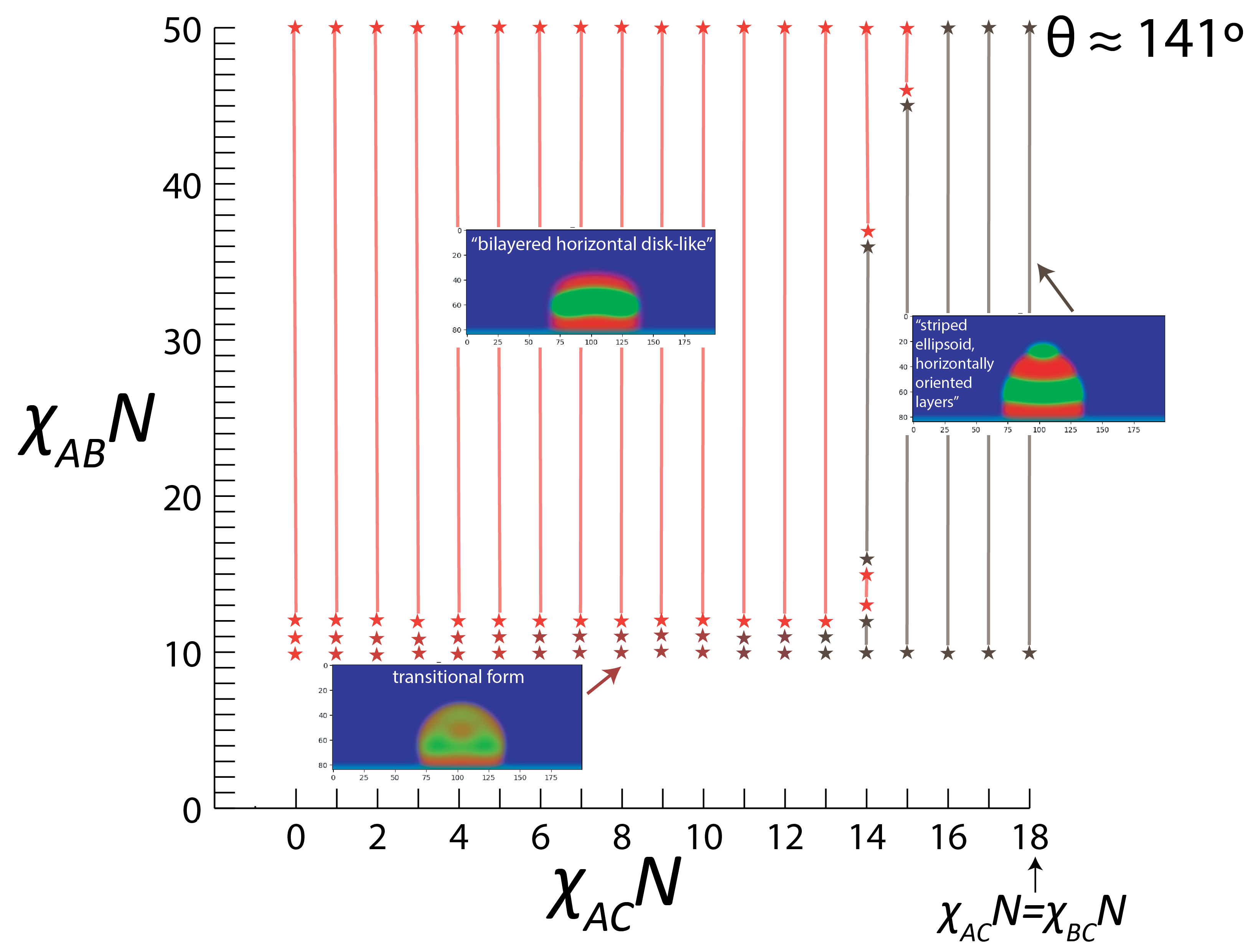}
	\caption{}
	\end{subfigure}
	\begin{subfigure}{0.49\textwidth}
	\includegraphics[width=\linewidth,height=\textheight,keepaspectratio]{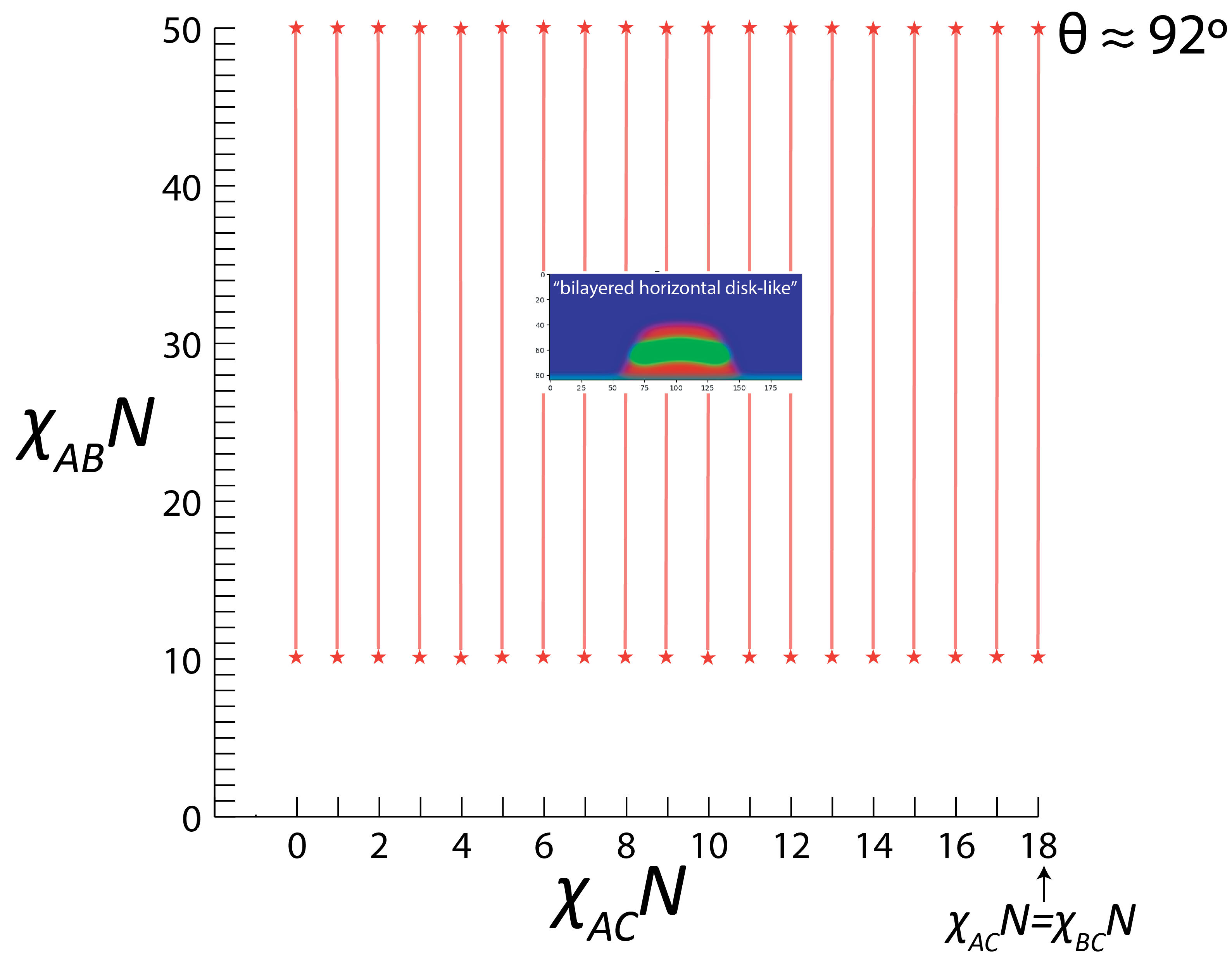}
	\caption{}
	\end{subfigure}
	\begin{subfigure}{0.49\textwidth}
	\includegraphics[width=\linewidth,height=\textheight,keepaspectratio]{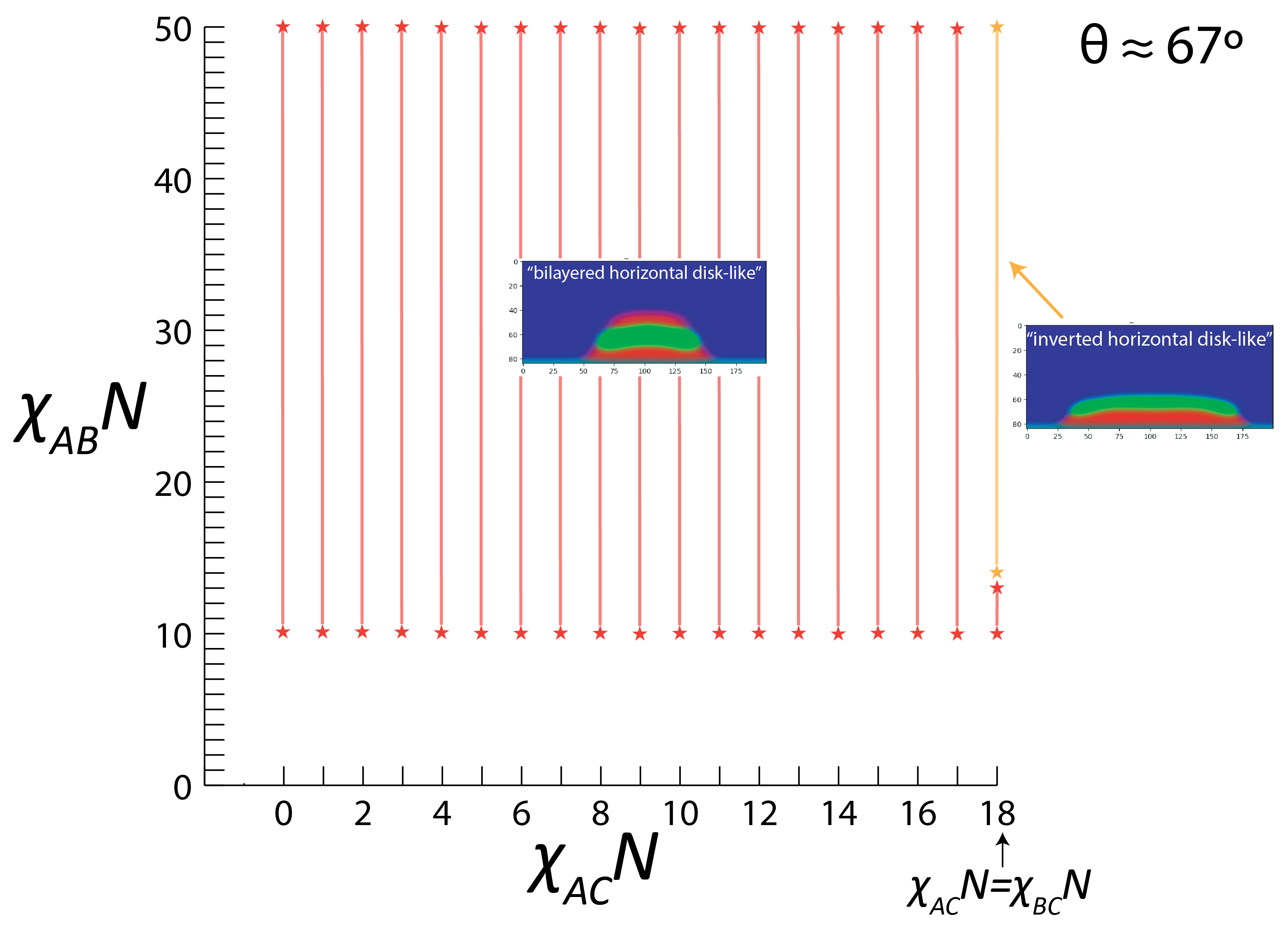}
	\caption{}
	\end{subfigure}
  \caption{Morphological diagram of nanodroplets adsorbed at a surface with "strong" preference towards $A$ blocks ($Pref_A\approx 0.157$); the surrounding medium is either neutral or preferential to the same block ($\chi_{AC}\leq\chi_{BC}$). SCFT calculations were performed in 2D space. Diagrams (a), (b), and (c) were obtained at $\theta\approx 141\degree$, $\theta\approx 92\degree$, and $\theta\approx 67\degree$, respectively. Points of different colors represent the boundary structures of different morphologies at a fixed $\chi_{AC}N$. Continuous lines between the boundary points show that this morphology was obtained at all values of $\chi_{AB}N$ between these points. For $\theta\approx141\degree$ and small $\chi_{AB}N\approx 10$, the striped ellipsoidal droplet transitioned into disk-like structure smoothly. We did not perform additional procedures to determine an exact boundary between these two morphologies in this region, since we did not carry out the iterative correction of this diagram. As a result, this transitional region is shown by points with color changing by a gradient between the color corresponding to the striped ellipsoidal droplet and the color corresponding to the disk-like structure. A characteristic snapshot for each structure as well as its short name are shown for each morphological region. $A$ blocks are shown in red, $B$ blocks are colored green, the surface-tethered $S$ homopolymers are shown in light blue, and homopolymers $C$ modeling the surrounding medium are colored dark blue.}
\label{fig:2}
\end{figure}

We also studied the morphologies of symmetric diblock copolymer nanodroplets (of the same size) adsorbed on surfaces preferring one type of segments over the other ($Pref_A>0$ or $Pref_A<0$). For definiteness, the surrounding medium was either neutral or preferential to $A$ block in this section ($\chi_{AC}\leq\chi_{BC}$); since the studied diblock copolymers are symmetric, this choice is arbitrary. Figure \ref{fig:2} shows the morphological diagram for strongly $A$-preferential surfaces ($Pref_A\approx0.157$) obtained using real-space SCFT calculations \cite{sides2003parallel,drolet1999combinatorial}. We did not use the iterative correction algorithm to obtain the equilibrium morphological boundaries in this diagram. However, we confirmed that at least one structure in each morphological region was in true equilibrium; therefore, the only difference between a true equilibrium diagram for $Pref_A\approx0.157$ and Figure \ref{fig:2} is the position of morphological boundaries and not the structures themselves (since it is very rare that a new equilibrium structure is produced during the ICA, compare Figure S2 and \ref{fig:1}).

Figure \ref{fig:2} shows that the nanodroplet structures that were metastable on a neutral surface (bilayered horizontal disk-like droplets and striped ellipsoids with horizontally oriented layers, see Figure S2) were the dominant equilibrium structures on $A$-preferential surfaces. This stabilization occurred since the $A$ blocks were strongly driven to form a surface-adsorbed layer, which forced other blocks to form layers parallel to the surface as well. Similar terrace-like structures were formed in experiments on much larger microdroplets of symmetric diblock copolymers \cite{croll2009spreading,croll2006droplet}. The striped ellipsoidal droplet with horizontally oriented layers was stabilized primarily when BCP wetted the surface poorly (Figure \ref{fig:2}a). Upon an increase of strength of BCP-surface wetting (Figure \ref{fig:2}b,c), the droplets tended to avoid a vertically-elongated shape necessary to stabilize the striped ellipsoidal structure. As a result, the bilayered disk-like droplet occupied the diagram at $\theta\leq90\degree$ almost entirely. When BCP wetted the surface strongly ($\theta\approx67\degree$, Figure \ref{fig:2}c), the droplets tended to flatten so much that a disk-like nanodroplet similar to the one formed on a neutral surface was formed (Figure \ref{fig:1}c) but with inverted layer ordering ($B$ block facing the surrounding medium, see Figure \ref{fig:2}c). Such structures were formed when $\chi_{AC}\approx\chi_{BC}$ due to a lower penalty to form a medium-exposed $B$ layer compared to the case of strongly $A$-preferential medium $\chi_{AC}<\chi_{BC}$. Such disk-like droplets were obtained experimentally in similar wetting and surface-to-block preference conditions \cite{hur2015interplay} (also see next section).

Importantly, when the surface preferred $A$ segments less strongly ($Pref_A\approx0.048$ and $Pref_A\approx0.094$, see Figure S5 and S6, respectively), the diagrams mostly featured the structures formed on strongly $A$-preferential surfaces (Figure \ref{fig:2}) with an addition of onion-like morphology obtained on neutral substrates (Figure \ref{fig:1}). Interestingly, there was one additional structure observed at intermediate $Pref_A$ that was not observed neither in Figure \ref{fig:1} nor in Figure \ref{fig:2}. This structure formed at weakly $A$-preferential surface ($Pref_A\approx0.048$, Figure S5c) and is a modified onion-like structure with segmented $A$ domains. This structure was metastable on neutral surfaces (Figure S2c); however, it may be equilibrium for weakly $A$-preferential surfaces due to additional energetic bonus for blocks $A$ to form surface-adsorbed domains.

Finally, when the surface was preferential to $B$ blocks ($Pref_A\approx -0.064$, Figure S7), we observed structures formed on non-preferential surfaces (Figure \ref{fig:1}) such as onion- and disk-like droplets, morphologies that desorbed from the surface, as well as film- and adsorbed micelle-like states. Interestingly, droplets on $B$-preferential surfaces also formed structures typical for $A$-preferential surfaces such as bilayered disk-like and the striped ellipsoidal droplets with horizontal layer orientation. However, in the case of $B$-preferential surface, the spatial positions of $A$ and $B$ blocks in these states were swapped compared to the structures formed on $A$-preferential substrates due to the preference of the surface to $B$ blocks instead of $A$ blocks.

It is worth mentioning that all structures obtained after SCFT calculations with nonzero $Pref_A$ were present after initial SCFT calculations with $Pref_A\approx0$ (Figure S2). Therefore, the calculations at $Pref_A\neq0$ could not supply any additional structures to the $Pref_A\approx0$ morphological diagram. Based on this observation, we made an assumption that we can view diagrams at each $Pref_A$ separately from each other, and no morphologies obtained at other values of $Pref_A$ and not at the given $Pref_A$ can be stable at this given $Pref_A$. In this approximation, the diagrams presented in Figure \ref{fig:1} and \ref{fig:2} present almost a full spectrum of characteristic structures (except for segmented onion-like structure observed in Figure S5c) that can be formed by symmetric diblock copolymers in small surface-adsorbed nanodroplets (with size on the order of $10-100$ nm).

In other words, there are two very distinct conditions driving significantly different self-assembly behavior: (i) placing a BCP droplet on a surface neutral to both blocks (Figure \ref{fig:1}) and (ii) placing a BCP droplet onto a substrate with a strong preference to one of the blocks and into a medium having preference to that same block (Figure \ref{fig:2}). Other self-assembly conditions do not seem to produce qualitatively different morphologies (Figure S5-S7).

As a result, there are approximately eight characteristic classes of structures that can be formed by a symmetric diblock copolymer in equilibrium when adsorbed on a surface (Figure \ref{fig:1}, \ref{fig:2}, S5-S7). These structures include striped ellipsoidal droplets with various number of vertically or horizontally-oriented layers, vertical, horizontal, and bilayered horizontal disk-like structures, segmented and ordinary onion-like droplets, and the adsorbed micelle-like state. In each of the structures, inversion of the block positioning is possible as well (compare Figure S7 with Figure \ref{fig:2}). Despite the two-dimensional nature of SCFT structures presented in this section, all these states can be reproduced in three dimensions, and some of these structures were obtained in experiments (see ref. \cite{hur2015interplay} and the next section). Moreover, these structures can be switched into each other by varying the interaction parameters, and this provides a way to realize this switch by adding surfactants and \textit{via} temperature or pH control (see next section). In addition, these eight characteristic structure classes were obtained by varying four independent parameters ($Pref_A$, $\theta$, $\chi_{AB}N$, and $\chi_{AC}N$). Surprisingly, this number of states is similar to the number of characteristic structures (six) formed by the same BCP in a nanodroplet of the same volume in solution, where only two independent parameters were varied ($\chi_{AB}N$ and $\chi_{AC}N$) \cite{petrov2025symmetric}. This demonstrates that making confinement more complex (\textit{i.e.}, introducing additional elements (such as surface) giving rise to a bigger number of independently varied system parameters) does not always lead to a significantly higher overall versatility of equilibrium BCP assemblies in this confinement.

\subsection{3D Space SCFT Calculations and Experimental Verification}

In this section, we studied the 3D equilibrium structure of the droplets adsorbed on a non-preferential surface ($Pref_A\approx 0$). We did not perform the 3D space SCFT calculations of non-adsorbed nanodroplets (since it was studied in our previous work \cite{petrov2025symmetric}); we also excluded the film-like and adsorbed micelle-like structures  obtained in Figure \ref{fig:1}b,c. As described in SI section 2, we created 3D analogs of the 2D equilibrium nanodroplet structures (taken from Figure \ref{fig:1}) and established their equilibrium position on a morphological diagram. Since this diagram was much coarser than the diagram of 2D nanodroplets (Figure \ref{fig:1}, \ref{fig:2}, S5-S7), we established the equilibrium position of each morphology by selecting the lowest free energy structure after real-space SCFT calculations initialized from all other structures in the diagram (SI section 2).

\begin{figure}[h!]
\centering
  \begin{subfigure}{0.49\textwidth}
	\includegraphics[width=\linewidth,height=\textheight,keepaspectratio]{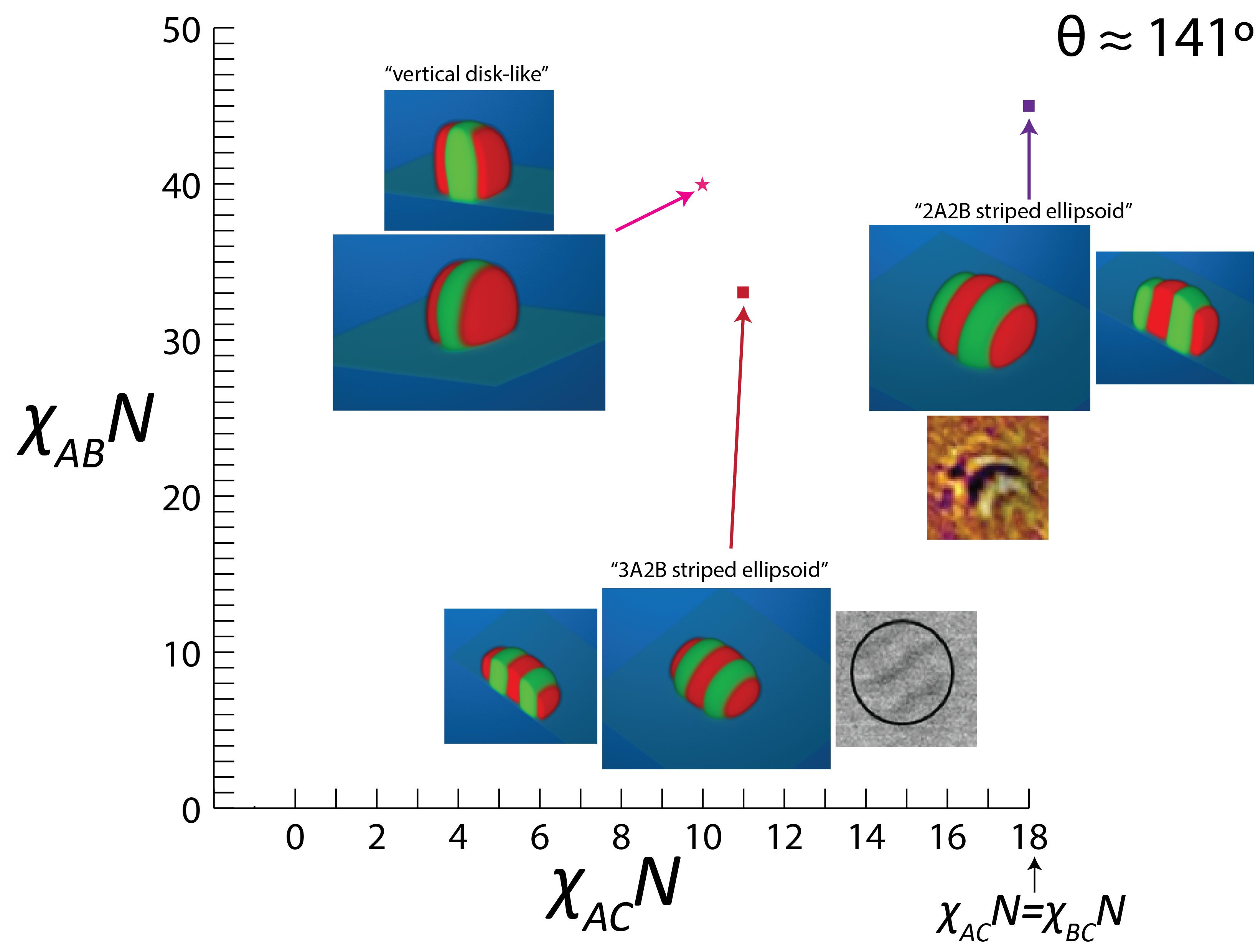}
	\caption{}
	\end{subfigure}
	\begin{subfigure}{0.49\textwidth}
	\includegraphics[width=\linewidth,height=\textheight,keepaspectratio]{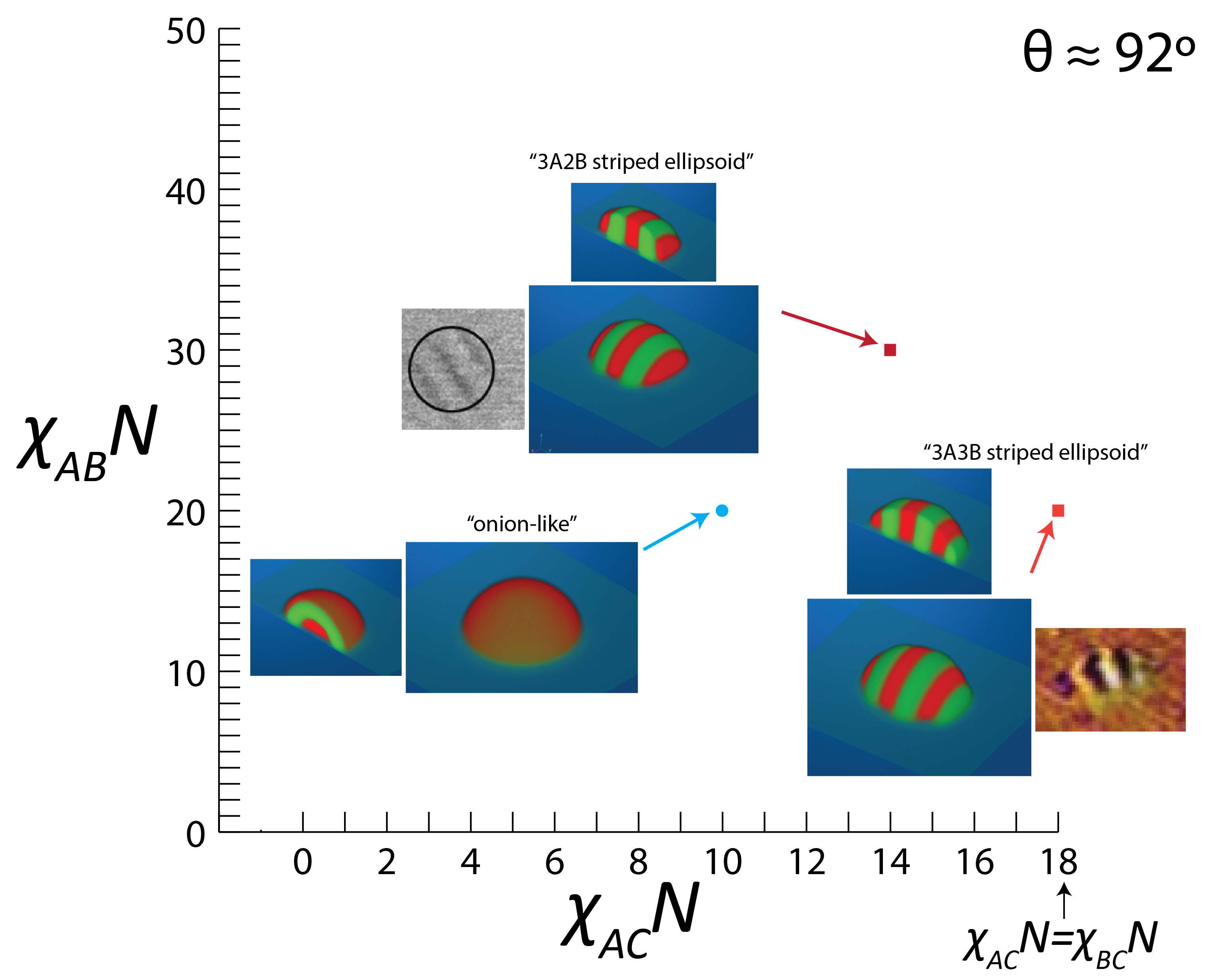}
	\caption{}
	\end{subfigure}
	\begin{subfigure}{0.49\textwidth}
	\includegraphics[width=\linewidth,height=\textheight,keepaspectratio]{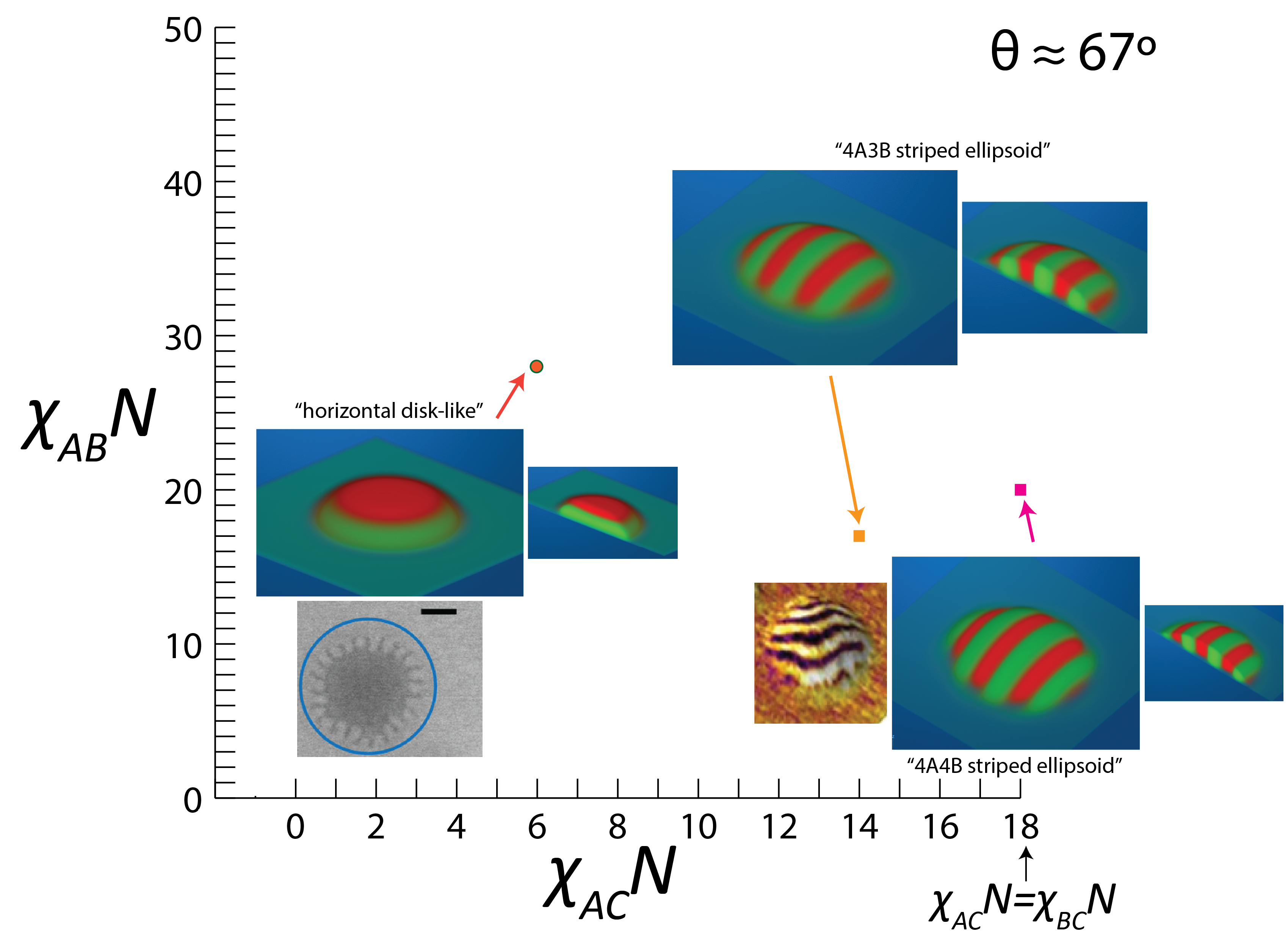}
	\caption{}
	\end{subfigure}
  \caption{Equilibrium morphological diagram of three-dimensional nanodroplets adsorbed at a surface with no preference towards any of the blocks ($Pref_A\approx 0$). Diagrams (a), (b), and (c) were obtained at $\theta\approx 141\degree$, $\theta\approx 92\degree$, and $\theta\approx 67\degree$, respectively. Points of different colors represent different morphologies. A characteristic snapshot for each structure as well as its short name are shown for each morphology. Smaller images located next to each structure show a slice of the morphology in the middle. $A$ blocks are shown in red, $B$ blocks are colored green, the surface-tethered $S$ homopolymers are shown in light cyan, and homopolymers $C$ modeling the surrounding medium are not shown. Experimental scanning electron microscopy and atomic force microscopy images were reproduced with permission from ref. \cite{hur2015interplay}. Copyright 2015 American Chemical Society. In (c), the scale bar is 100 nm.}
\label{fig:3}
\end{figure}

This equilibrium morphological diagram is shown in Figure \ref{fig:3}. First, the comparison of Figure \ref{fig:1} and \ref{fig:3} shows that the vast majority of discovered nanodroplet structures in 2D space were reproduced in 3D space, and their equilibrium positions were very similar in both diagrams. Therefore, the conclusions about the nanodroplet phase behavior drawn from the 2D space calculations in the previous section should be qualitatively accurate when applied to real three-dimensional nanodroplets. Moreover, Figure \ref{fig:3} demonstrates the experimental confirmation of the existence of the majority of discovered morphologies. In particular, the striped ellipsoidal nanodroplets with various number of lamellar layers similar to those obtained in SCFT calculations were obtained experimentally \cite{hur2015interplay} on a neutral surface. This demonstrates the feasibility of experimental preparation of small nanodroplets with size on the order of tens of nanometers with well-tunable nanostructure. Moreover, the horizontal disk-like droplet discovered in 2D (Figure \ref{fig:1}c, \ref{fig:2}c) and 3D (Figure \ref{fig:3}c) space SCFT calculations was obtained in experiment \cite{hur2015interplay} (Figure \ref{fig:3}c). Due to a much larger size, the experimental disk structure had edges with vertically oriented lamellae; however, the bulk of the experimentally observed morphology was identical to the smaller horizontal disk-like droplet obtained in SCFT. As a result, the calculations presented in this work can be an accurate qualitative tool to guide experiments on the hierarchical droplet-based materials.

Interestingly, several equilibrium structures obtained in this work were not observed in the previous available experimental study of small BCP nanodroplets \cite{hur2015interplay}. These structures include the bilayered vertically oriented disk-like nanodroplet, the onion-like structure (Figure \ref{fig:3}), and the striped ellipsoidal droplet with horizontally oriented layers (Figure \ref{fig:2}a). This could be explained by the predominant existence of these structures at rather large $\theta\geq 90\degree$ in equilibrium (see Figure \ref{fig:1}a,b, \ref{fig:3}a,b, S5a, S6a, and S7a,b). On the other hand, the droplet profiles obtained using atomic force microscopy in ref. \cite{hur2015interplay} (Figure \ref{fig:3}) showed them as wetting the surface rather well. This suggested rather small $\theta$ accessed in ref. \cite{hur2015interplay}, possibly insufficient to obtain the newly predicted nanodroplets described above.

We posit that such structures can be discovered experimentally by decreasing the BCP-surface wetting strength \textit{via} the addition of surfactants that increase the compatibility of the surface (typically crosslinked random copolymers with different composition that controls the preference to one of the blocks \cite{hur2015interplay}) and the surrounding medium (typically gas or vapor). We designed the following computational experiment to show that this route may be experimentally viable. One can obtain amphiphilic surfactant molecules that readily adsorb on the surface and have selective preference to the surrounding medium higher than the bare surface. We modeled such molecules as short $DE$ symmetric diblock copolymers (10 times shorter than the $AB$ diblock copolymers that formed the droplet). Both $D$ and $E$ blocks were fully compatible with the surface ($\chi_{DS}=\chi_{ES}=0$), and blocks $E$ were fully compatible with the surrounding medium ($\chi_{CE}=0$). On the other hand, blocks $D$ were strongly incompatible with the medium ($\chi_{CD}N=100$). Both $D$ and $E$ blocks were taken to be incompatible with $A$ and $B$ segments to avoid the adsorption of the "surfactants" onto the droplet; however, this was done just for the simplicity of the proof of concept and should not be generally the case. In fact, surfactants can be used to modulate the preference of one of the blocks to the surrounding medium by adsorbing on a droplet and \textit{via} subsequent light irradiation or temperature control (as done in experiments on block copolymer nanoparticles in solution \cite{hu2021light}).

\begin{figure}
  \includegraphics[width=\linewidth]{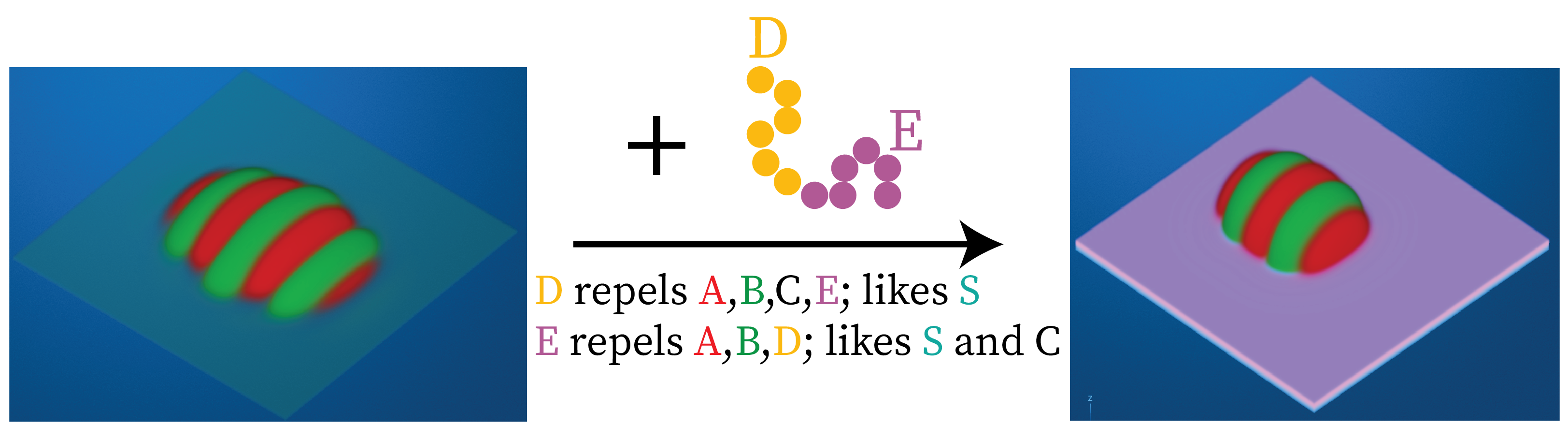}
  \caption{The change of equilibrium nanodroplet structure upon addition of short $DE$ symmetric diblock copolymers that modeled amphiphilic surfactant molecules. $A$ and $B$ blocks are shown in red and green. Before the addition of surfactant, the surface only consisted of $S$ homopolymers modeling the crosslinked random copolymer substrates used in experiment \cite{hur2015interplay} (light cyan). $S$ was moderately incompatible with the surrounding medium modeled by homopolymer $C$ (not shown). After the addition, $DE$ diblock copolymers adsorbed at the surface forming a three-layer substrate (blocks $D$ are shown in light orange, and blocks $E$ are colored in violet). The blocks $E$ are located at the top of the surface due to their high compatibility with the surrounding medium. This addition led to an increase of effective $\theta$.}
  \label{fig:surfactant}
\end{figure}

Figure \ref{fig:surfactant} shows that the adsorption of such "surfactant" molecules onto the surface led to a strong increase of the effective $\theta$ value and to the transition between different equilibrium droplet nanostructures (from $4A3B$ to $3A2B$ striped ellipsoidal droplet). Thus, one can utilize surfactants to access large-$\theta$ conditions in experiments and obtain the previously undiscovered structures predicted in Figure \ref{fig:2} and \ref{fig:3}. On a broader scale, one can effectively tune the $Pref_A$, $\theta$, $\chi_{AC}N$, and $\chi_{AB}N$ parameters by using surfactant molecules; this provides a leverage to externally control the droplet morphology and switch between the nanostructures obtained in Figure \ref{fig:1}-\ref{fig:3}, S5-S7.

To conclude, Figure \ref{fig:1}-\ref{fig:surfactant} demonstrate that it is possible to switch between nanodroplets having various shape and the chemical composition exposed to the medium. This can be realized by tuning the interaction parameters between the system components. The presented data also reveals the rules according to which the parameters should be changed in order to transition between different types of nanodroplets. In turn, the reversible switch between the droplets exhibiting predominantly $A$ blocks to the surrounding medium (like onion-like or horizontal disk-like droplets) and the droplets exhibiting roughly equal amount of $A$ and $B$ blocks (such as striped ellipsoids) will lead to a change in chemical composition of droplet-coated surfaces. Moreover, Figure \ref{fig:3} predicts the existence of reversible switching between anisotropic particles such as striped ellipsoids or vertically-oriented disks to isotropic nanostructures such as onion-like droplets and horizontally-oriented disks. As a result, the topographical nanostructure and optical anisotropy of BCP droplet-coated surfaces can be externally tuned as well. This switching can be realized experimentally, as described in the previous paragraph, \textit{via} addition of surfactants and by light irradiation or temperature/pH control. Notably, our calculations supported by existing experimental data suggest that such tuning of surface chemistry and topography can be realized using the simplest symmetric diblock copolymers.

\section{Conclusion}

In this work, we performed a comprehensive investigation of the phase behavior of symmetric diblock copolymer surface-adsorbed nanodroplets. Augmented with the previously published experimental data \cite{hur2015interplay} as well as with the simulation of amphiphilic molecule addition (Figure \ref{fig:surfactant}), we demonstrated that the studied small ($10-100$ nm) nanodroplets form highly versatile nanostructures, between which direct transitions are possible by applying external stimuli. As a result, such small droplets can serve as the building blocks for hierarchically organized surface coatings with externally switchable nanostructure. We discovered eight classes of morphologies that the symmetric diblock copolymer nanodroplets can form in equilibrium by varying the available interaction parameters in the system as well as by using a newly developed iterative correction algorithm. The latter produced equilibrium morphological diagrams without the need for guessing the potential equilibrium structures beforehand. We also discovered that making the confinement more complex (\textit{i.e.}, adsorbing block copolymer droplets in solution onto a surface, which increases the number of control parameters) does not always lead to a significant rise of the overall structural complexity of equilibrium assemblies. This finding points to an unexpected "saturation of complexity" of confined block copolymer phase behavior, the physical roots of which are a good topic for future research. We predicted how to obtain all observed structures experimentally and demonstrated that the simplest type of block copolymers can serve as a platform for engineering stimuli-responsive hierarchically structured surface coatings for a variety of industrial applications.

\section{Methods}

The structure of surface-adsorbed nanodroplets of symmetric diblock copolymers was modeled using the self-consistent field theory (SCFT). In this section, we will describe the procedure of SCFT modeling in 2D space. For SCFT calculations in 3D space (section 2.2), we used similar techniques except for the procedure of the initial droplet preparation and except for the algorithm of obtaining the global free energy minimum structure (see SI section 2 for details). In 2D space modeling, for the droplets adsorbed on a non-preferential surface, SCFT was augmented with the newly developed iterative correction algorithm (ICA) in order to obtain droplet structures corresponding to the global minimum of SCFT free energy. Section 4.1 describes the technique of SCFT calculations. Section 4.2 describes the ICA and the protocol of verifying that ICA yielded near-equilibrium morphological diagram of nanodroplets adsorbed on neutral surfaces.

\subsection{SCFT Calculations}

In SCFT, one represents a real polymer chain with an infinitely thin thread with Gaussian statistics. In order to obtain the equilibrium nanodroplet structure in SCFT, one has to obtain the global minimum of the following functional of the mean-field excess free energy per chain:

\begin{equation}
\label{eq:freeen}
    f=\frac{1}{V}\int d\textbf{r} \left[\frac{1}{2}\sum_{i\neq j} \chi_{ij}N(\phi_i(\textbf{r})-\bar{\phi_i})(\phi_j(\textbf{r})-\bar{\phi_j}) - \sum_{i} \omega_i(\textbf{r}) \phi_i(\textbf{r}) - p(\textbf{r})(1-\sum_{i} \phi_i(\textbf{r})) - \text{ln}Q \right]
\end{equation}

In Equation \ref{eq:freeen}, $\chi_{ij}$ is the Flory-Huggins parameter that determines the energy of interactions between species $i$ and $j$. In SCFT, the reference length of chains $N$ is used to set the scale of distances (by determining the chain radius of gyration $R_g$) and the scale of chemical potential field for species $i$ ($\omega_i$); $N$ was fixed in our SCFT calculations. As traditionally done in SCFT BCP nanoparticle research \cite{petrov2025symmetric,kim2009droplets,huang2024design}, the medium surrounding a BCP droplet was modeled by homopolymers $C$ having the same $N$ as BCP chains; as ref. \cite{petrov2025symmetric} showed, this modeling approach did not affect the qualitative correctness of the model when applied to experiments. The $S$-homopolymer modeling a polymer-coated surface was shorter: $N_S=0.1N$. We fixed $\chi_{BC}N=18$ as in ref. \cite{petrov2025symmetric} in order to provide the soft-confinement conditions for the block copolymers, \textit{i.e.}, to allow the droplet to deform during the self-assembly but preclude the droplet from breaking apart at the majority of studied interaction parameters. Other $\chi_{ij}N$ parameters were varied as described in Results and Discussion; $\chi_{AB}N$ and $\chi_{AC}N$ were changed with a step $\Delta\chi_{AB}N=\Delta\chi_{AC}N=1$. As demonstrated in ref. \cite{petrov2025symmetric}, this step was small enough to find "rare" morphologies that existed in a small range of interaction parameters.

In Equation \ref{eq:freeen}, $\phi_i(\textbf{r})$ and $\bar{\phi_i}$ are the volume fractions of the species $i$ at the point $\textbf{r}$ and on average, respectively. $\phi_i(\textbf{r})$ describe the spatial distribution of species in the system and define the morphology of a BCP droplet. $p(\textbf{r})$ is the pressure field that enforces the incompressibility of the system (\textit{i.e.}, $\sum_{i} \phi_i(\textbf{r}) = 1$). Finally, $Q$ is the single chain partition function. For the systems studied in this work, $\text{ln}Q = (\bar{\phi_{A}}+\bar{\phi_{B}})\text{ln}Q_{DBC}+\bar{\phi_{C}}\text{ln}Q_{C,hom}+(1-\bar{\phi_{A}}-\bar{\phi_{B}}-\bar{\phi_{C}})\text{ln}Q_{S,hom}$, where $Q_{DBC}$, $Q_{C,hom}$, and $Q_{S,hom}$ are the partition functions of a diblock copolymer chain, a $C$-homopolymer, and an $S$-homopolymer tethered at the surface, respectively. $Q_{DBC}$, $Q_{C,hom}$, and $Q_{S,hom}$ can be derived from the restricted partition functions $q_{DBC}(\textbf{r},s)$, $q_{C,hom}(\textbf{r},s)$, and $q_{S,hom}(\textbf{r},s)$: $Q_{DBC}=V^{-1}\int d\textbf{r}q_{DBC}(\textbf{r},1)$, $Q_{C,hom}=V^{-1}\int d\textbf{r}q_{C,hom}(\textbf{r},1)$, and $Q_{S,hom}=V^{-1}\int d\textbf{r}q_{S,hom}(\textbf{r},0.1)$. The variable $s \in (0,1)$ parametrizes the position along the chain length $N$, $N_s=100$ steps were taken along the chain to resolve the propagator during the numerical solution. The restricted partition functions and their conjugates $q_{DBC}^{\dag}(\textbf{r},s)$, $q^{\dag}_{C,hom}(\textbf{r},s)$, and $q^{\dag}_{S,hom}(\textbf{r},s)$ are the solutions of the modified diffusion equations (Equation \ref{eq:qab}, \ref{eq:qc}, and \ref{eq:qs}) with the initial conditions $q_{DBC}(\textbf{r},0)=q_{C,hom}(\textbf{r},0)=q_{DBC}^{\dag}(\textbf{r},1)=q_{C,hom}^{\dag}(\textbf{r},1)=q_{S,hom}^{\dag}(\textbf{r},0.1)=1$. $q_{S,hom}(\textbf{r},0)=1$ for all $\textbf{r}$ located at the surface (\textit{i.e.}, at $z=0$, where $z$ is the axis perpendicular to the surface). $q_{S,hom}(\textbf{r},0)=0$ for all other $\textbf{r}$. Such initial conditions ensured tethering of short $S$ homopolymers at the surface.

\begin{equation}
\label{eq:qab}
\begin{cases}
    \frac{\partial q_{DBC}}{\partial s}= R_g^2 \nabla^2 q_{DBC}(\textbf{r},s) - \omega_A(\textbf{r}) q_{DBC}(\textbf{r},s), s\in(0,f) \\
    \frac{\partial q_{DBC}}{\partial s}= R_g^2 \nabla^2 q_{DBC}(\textbf{r},s) - \omega_B(\textbf{r}) q_{DBC}(\textbf{r},s), s\in(f,1) \\
    -\frac{\partial q^{\dag}_{DBC}}{\partial s}= R_g^2 \nabla^2 q^{\dag}_{DBC}(\textbf{r},s) - \omega_A(\textbf{r}) q^{\dag}_{DBC}(\textbf{r},s), s\in(0,f) \\
    -\frac{\partial q^{\dag}_{DBC}}{\partial s}= R_g^2 \nabla^2 q^{\dag}_{DBC}(\textbf{r},s) - \omega_B(\textbf{r}) q^{\dag}_{DBC}(\textbf{r},s), s\in(f,1)
    \end{cases}
\end{equation}

\begin{equation}
\label{eq:qc}
\begin{cases}
    \frac{\partial q_{C,hom}}{\partial s}= R_g^2 \nabla^2 q_{C,hom}(\textbf{r},s) - \omega_C(\textbf{r}) q_{C,hom}(\textbf{r},s), s\in(0,1) \\
    -\frac{\partial q^{\dag}_{C,hom}}{\partial s}= R_g^2 \nabla^2 q^{\dag}_{C,hom}(\textbf{r},s) - \omega_C(\textbf{r}) q^{\dag}_{C,hom}(\textbf{r},s), s\in(0,1)
    \end{cases}
\end{equation}

\begin{equation}
\label{eq:qs}
\begin{cases}
    \frac{\partial q_{S,hom}}{\partial s}= R_g^2 \nabla^2 q_{S,hom}(\textbf{r},s) - \omega_S(\textbf{r}) q_{S,hom}(\textbf{r},s), s\in(0,0.1) \\
    -\frac{\partial q^{\dag}_{S,hom}}{\partial s}= R_g^2 \nabla^2 q^{\dag}_{S,hom}(\textbf{r},s) - \omega_S(\textbf{r}) q^{\dag}_{S,hom}(\textbf{r},s), s\in(0,0.1)
    \end{cases}
\end{equation}

Here, $f=0.5$ is the fraction of A segments in a diblock copolymer. The saddle point of the free energy (Equation \ref{eq:freeen}) is attained when fields $\omega_i(\textbf{r})$, $\phi_i(\textbf{r})$, and $p(\textbf{r})$ are given by Equation \ref{eq:saddle}:

\begin{equation}
\label{eq:saddle}
\begin{cases}
    \omega_i(\textbf{r}) = \sum_{j\neq i} \chi_{ij}N(\phi_j(\textbf{r})-\bar{\phi_j})+p(\textbf{r}) \\
    \sum_{i} \phi_i(\textbf{r}) = 1 \\
    \phi_A(\textbf{r}) = (\bar{\phi_A}+\bar{\phi_B})Q_{DBC}^{-1} \int_{0}^{f} ds q_{DBC}(\textbf{r},s)q^{\dag}_{DBC}(\textbf{r},s) \\
    \phi_B(\textbf{r}) = (\bar{\phi_A}+\bar{\phi_B})Q_{DBC}^{-1} \int_{f}^{1} ds q_{DBC}(\textbf{r},s)q^{\dag}_{DBC}(\textbf{r},s) \\
    \phi_C(\textbf{r}) = \bar{\phi_C}Q_{C,hom}^{-1} \int_{0}^{1} ds q_{C,hom}(\textbf{r},s)q^{\dag}_{C,hom}(\textbf{r},s) \\
    \phi_S(\textbf{r}) = (1-\bar{\phi_A}-\bar{\phi_B}-\bar{\phi_C})Q_{S,hom}^{-1} \int_{0}^{0.1} ds q_{S,hom}(\textbf{r},s)q^{\dag}_{S,hom}(\textbf{r},s)
    \end{cases}
\end{equation}

We used the real-space algorithm \cite{sides2003parallel,drolet1999combinatorial} to solve the Equation \ref{eq:qab}, \ref{eq:qc}, \ref{eq:qs}, \ref{eq:saddle}. Lattice site resolution was fixed at $0.1R_g\times0.1R_g$. The simulation cell had dimensions $x\times z$: $20R_g\times8.4R_g$. Periodic boundary conditions were applied at $x=0$ and $x=20R_g$, Dirichlet boundary conditions were applied at $z=0$ and $z=8.4R_g$. At these values of $z$, propagators $q_{S,hom}$ (except for the first segment), $q_{DBC}$, and $q_{C,hom}$ were equal to zero to model the presence of surface at $z=0$. BCP volume fraction was equal to $0.178572$, homopolymer $S$ volume fraction was equal to $0.04$. The droplet area (the analog of volume in 3D space SCFT calculations) was fixed to $A_d=30R_g^2$. Therefore, the characteristic droplet size $\propto A_d^{1/2}$ was on the order of tens of nanometers for typical BCPs (with Kuhn length $\approx 1$ nm and degree of polymerization $\propto 10^2$). $A_d$ used in this work was equal to $A_d$ set in our previous study \cite{petrov2025symmetric} for the "ultrasmall" droplets in solution. In turn, this allowed us to compare the versatility of phase behavior of the same symmetric diblock copolymer droplet in two different soft confinement settings.

The calculations proceeded as follows: first, we created an equilibrium surface-adsorbed structureless droplet ($\chi_{AB}=0$) at target $\chi_{AC}N$, $\theta$, and $Pref_A$. Second, we set $\chi_{AB}$ to a target value and performed SCFT calculations until the free energy (Equation \ref{eq:freeen}) stabilized (\textit{i.e.}, the relative change of free energy in the last $5\times 10^4$ iterations was less than $5\times 10^{-6}$). This energy threshold was small enough such that all different morphologies generated in this work had energy differences exceeding this threshold when calculated at the same set of interaction parameters. At the same time, this threshold was large enough to make the calculations computationally feasible. As a result, by varying $\chi_{AB}N$, $\chi_{AC}N$, $\theta$, and $Pref_A$ in the ranges described in Results and Discussion, we created an "initial" four-dimensional morphological diagram of nanodroplets. SI section 3 describes the definition of a "morphology" and the algorithm to divide the diagram into morphological regions.

Importantly, not all structures in this "initial" diagram were at equilibrium at the set of $\chi_{AB}N$, $\chi_{AC}N$, $\theta$, and $Pref_A$ parameters they were obtained. The complex free energy landscape of block copolymer nanodroplets adsorbed on a surface gave rise to the local minima of the free energy (Equation \ref{eq:freeen}), which led to an abundant formation of metastable states. In the following section, we describe the procedure that turned this "initial" SCFT diagram of nanodroplet morphologies into near-equilibrium "final" diagram for the case of $Pref_A\approx0$.

\subsection{Iterative Correction Algorithm (ICA)}

In this section, we will describe the algorithm designed to obtain the global minimum (equilibrium) solution of SCFT equations at a given set of interaction parameters. ICA is a development of the older, simpler algorithm designed in our previous work \cite{petrov2025symmetric}. The main assumption underlying the algorithm is that for a given set of interaction parameters, the equilibrium droplet nanostructure can be obtained \textit{via} the real-space SCFT calculations either (i) at these interaction parameters or (ii) at some other set of interaction parameters. The assumption (ii) implies that if one varies the interaction parameters in a wide enough range, one will be able to obtain the entire set of equilibrium structures for this range after real-space SCFT calculations. However, these structures are not guaranteed to appear at exactly the interaction parameters at which they are equilibrium. Therefore, a certain algorithm that "reshuffles" the results of real-space SCFT calculations across the studied range of interaction parameters is required to obtain the equilibrium morphological diagram of nanodroplets. In principle, such algorithm would require the SCFT calculation of the "initial" diagram and subsequent calculations of all obtained droplet morphologies at all studied interaction parameters. However, since the number of independent parameters and their range are large, the number of obtained structures is prohibitively big to perform such "brute force" reshuffling procedure. Moreover, at each iteration of such simulations, new metastable structures will form, which will further complicate the simulations.

An efficient algorithm can be developed by making a mild assumption about the nature of the "initial" diagram: besides containing all equilibrium structures for a given range of varied parameters, at least one structure can be assumed to be at true equilibrium in the "initial" diagram in each simply connected region of each of these equilibrium structures. This additional assumption allowed us to perform the "reshuffling" of structures computationally efficiently to reach the true equilibrium diagram.

\begin{figure}[h!]
  \includegraphics[width=\linewidth]{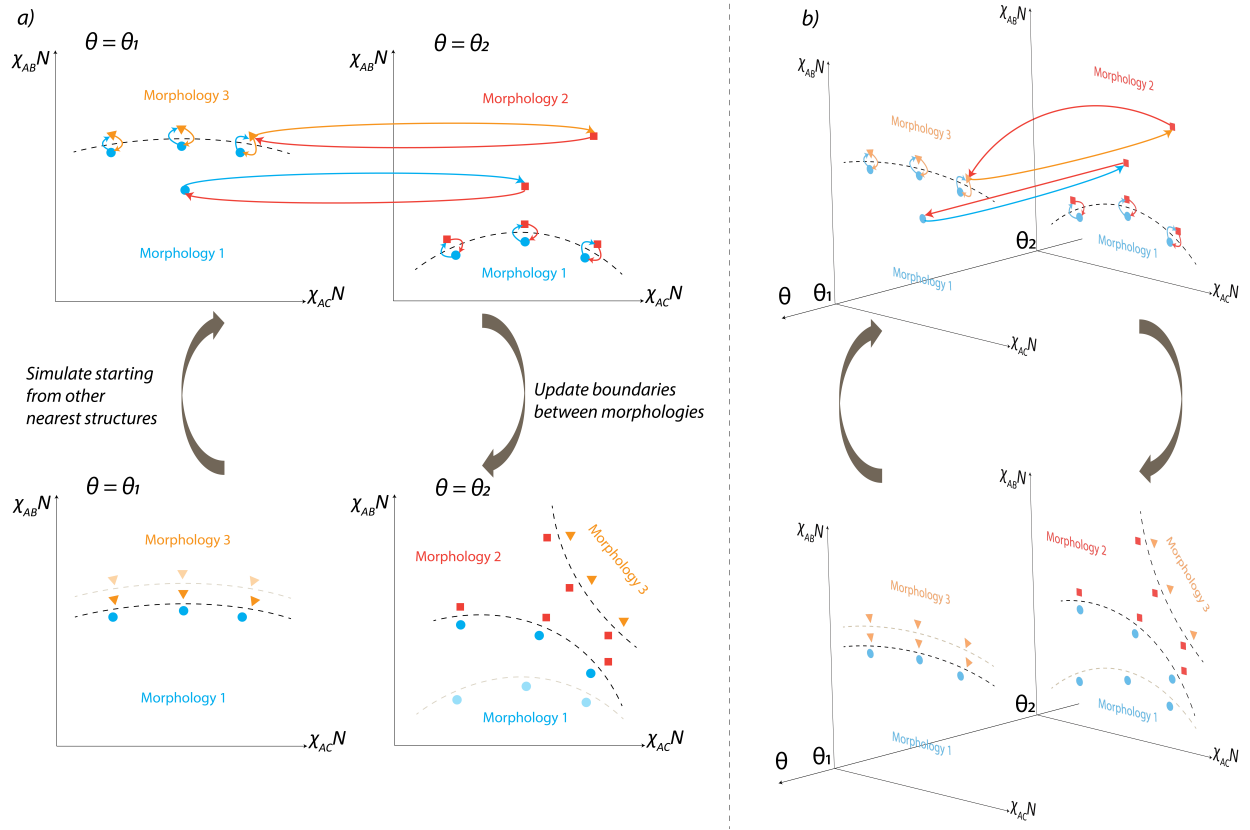}
  \caption{Schematic depiction of the iterative correction algorithm (ICA). For clarity, we showed a diagram featuring only two different values of $\theta$. (a) Each 2D diagram is a slice of the 3D diagram at a fixed $\theta$. Dashed lines show boundaries between morphologies; symbols denote example structures lying on the borders between morphologies. Arrows denote SCFT calculations in which a structure at the base of the arrow is taken as an initial structure for calculation at interaction parameters set as at the tip of the arrow. Long colored arrows depict calculations done between the planes with different values of $\theta$; only two pairs of such calculations are shown for visual clarity. After calculations are done (top), the morphological boundaries are updated by choosing the lowest-energy structure (bottom). Morphological boundaries and border structures before the update are shown in faint colors; fully-colored lines and symbols depict the boundaries and border structures, respectively, after the update. The calculations are performed again in the updated diagram according to the same algorithm until the borders stabilize. (b) Depiction of the same algorithm in a 3D diagram.}
  \label{fig:phdiagupdate}
\end{figure}

First, we generated the "initial" diagram of states for $Pref_A\approx0$ as described in the previous section (Figure S2). Second, we selected a structure lying at the border of a morphological region obtained at certain interaction parameters $\chi_{AB}N$, $\chi_{AC}N$, $\theta$. Third, we took nearby border structures belonging to other morphological regions and used them as initial states for SCFT real-space calculations at the interaction parameters $\chi_{AB}N$, $\chi_{AC}N$, $\theta$. In the implementation of the algorithm in this work, the nearby structures included six states obtained at the following interaction parameters: (i) $\chi_{AB}N\pm1$, $\chi_{AC}N$, $\theta$, (ii) $\chi_{AB}N$, $\chi_{AC}N\pm1$, $\theta$, and (iii) $\chi_{AB}N$, $\chi_{AC}N$, with any other value of $\theta$. Finally, after these calculations, we compared the free energies of the obtained structures and replaced the initial morphology with the lowest energy state. Such simulations were performed for all border structures in the "initial" diagram; after this, the diagram was updated (see Figure S3 as an example of such update). After performing these steps iteratively, the "initial" diagram of droplet morphologies was guaranteed to converge to a true equilibrium diagram (if the aforementioned assumptions about the "initial" diagram hold true). Figure \ref{fig:phdiagupdate} includes the scheme of this iterative correction algorithm (ICA). It is worth mentioning that during the ICA, we ignored the regions where the equilibrium structures were non-adsorbed droplets (see Figure \ref{fig:1}a); such structures required the SCFT calculations in a larger simulation cell and, therefore, their equilibrium regions were established according to a separate procedure (see SI section 4).

However, performing this algorithm sometimes led to a complete "migration" of a (simply connected) morphological region along the diagram (therefore, not a single state in this region in the "initial" diagram was in true equilibrium, compare Figure S2 and \ref{fig:1}). Moreover, one structure in the "final" diagram appeared not in the "initial" diagram but after the boundary update calculations. This, in turn, meant that the assumptions that guarantee the convergence of ICA to a true equilibrium diagram did not always hold. However, these assumptions are a sufficient but not a necessary condition for the applicability of this algorithm. To support the close-to-equilibrium nature of the "final" diagram, we performed the following test. We took each morphology obtained either in the "initial" diagram or during the ICA calculations as an initial state for real-space SCFT calculations at the interaction parameters approximately corresponding to the center of every morphological region in the "final" diagram (Figure \ref{fig:1}). After these calculations, we compared free energies of the structures and selected the lowest-energy one. We confirmed that these lowest-energy structures always coincided with the structures in Figure \ref{fig:1}. As a result, assuming that all possible equilibrium structures were obtained either in the "initial" diagram or during the ICA, we proved that the ICA yielded true equilibrium structures at the centers of all morphological regions in the "final" diagram. Therefore, despite the (rare) violations of the sufficient condition guaranteeing the convergence of the ICA to an equilibrium diagram, this algorithm was nevertheless able to yield the close-to-equilibrium diagram of nanodroplet structures. Establishing the rigorous necessary condition for the applicability of ICA is a good topic for future research.

\section*{Acknowledgements}

We acknowledge the MIT SuperCloud and Lincoln Laboratory Supercomputing Center as well as the MIT Office of Research Computing and Data for providing high performance computing resources. This research was supported by the National Science Foundation through award number DMREF 2118678.

\section*{Supporting Information}

Additional details of SCFT calculation protocols, morphological diagrams obtained during the iterative correction algorithm, morphological diagrams for surfaces preferring B blocks and for surfaces with weak and moderate preference to A blocks, and other data supporting claims in the main text.

\clearpage
\setcounter{page}{1}

\section*{Supporting Information}
\addcontentsline{toc}{section}{Supporting Information}

\setcounter{figure}{0}
\setcounter{table}{0}
\setcounter{equation}{0}
\setcounter{section}{0}

\renewcommand{\thefigure}{S\arabic{figure}}
\renewcommand{\thetable}{S\arabic{table}}
\renewcommand{\theequation}{S\arabic{equation}}
\renewcommand{\thesection}{S\arabic{section}}

\section{Mapping Between $\chi_{ij}N$ Values and $\theta$, $Pref_A$ Quantities}

As outlined in Results and Discussion, we considered four essential physical parameters that determined the morphology of a droplet in this work: $\chi_{AB}N$, $\chi_{AC}N$, $\theta$, and $Pref_A$ (see Results and Discussion for details). In this section, we will outline the exact definition of $\theta$ and $Pref_A$ and how we controlled these quantities using the $\chi_{ij}N$ parameters set in an SCFT calculation.

\begin{figure}
\centering
  \includegraphics[width=0.6\linewidth]{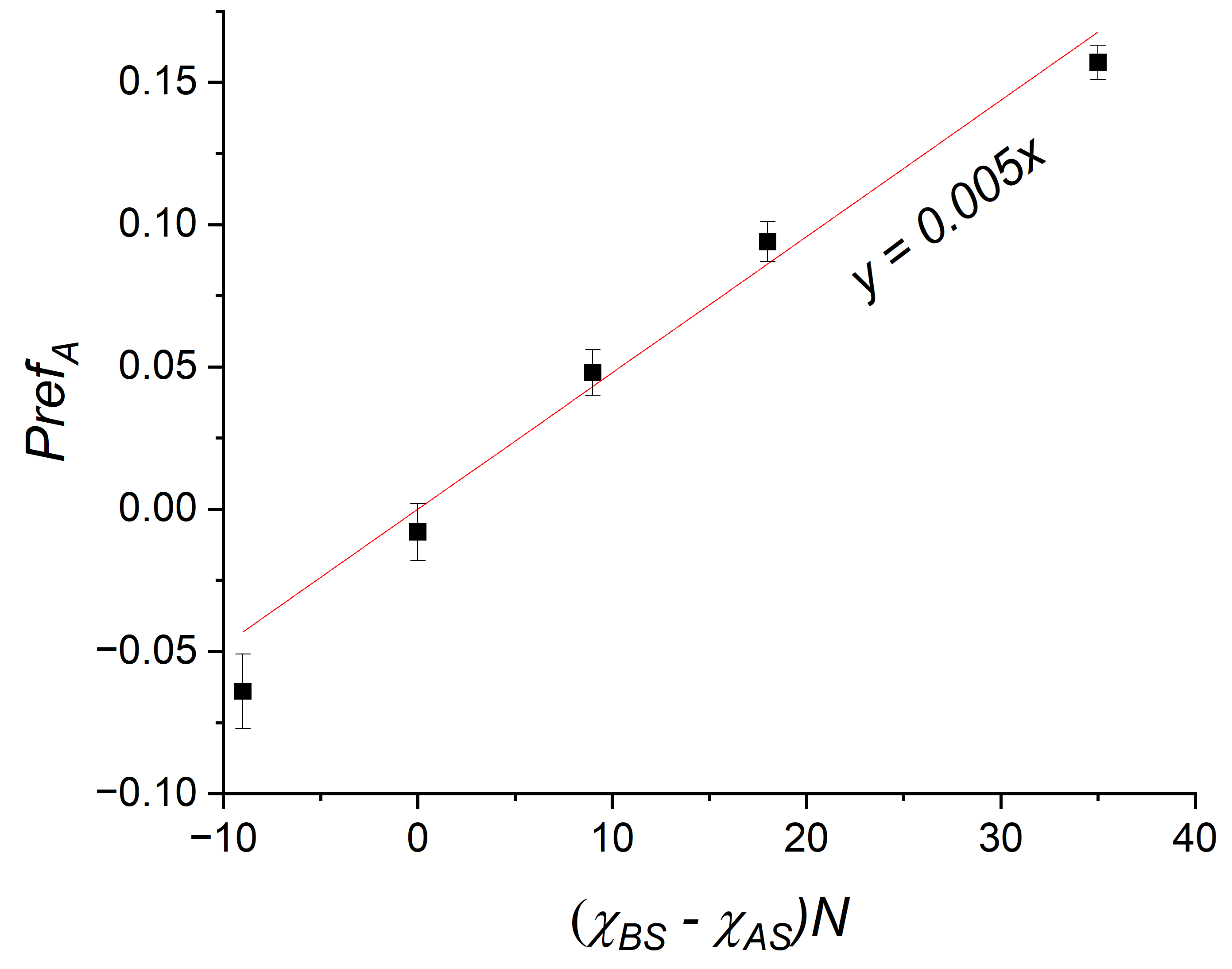}
  \caption{The dependency of $Pref_A$ (defined above) on the difference of surface-block repulsion parameters: $\chi_{BS}N-\chi_{AS}N$. Red line shows the best linear fit. Error bars show the standard deviation of $Pref_A$ measured in droplets with different $\theta$ and $\chi_{AC}N$.}
  \label{fig_s_prefa}
\end{figure}

First, let us precisely define $Pref_A$. As outlined in Results and Discussion, $Pref_A=\phi_A^{surf}-0.5$, where $\phi_A^{surf}$ is the fraction of $A$-type segments in a $z\in[0,0.9R_g]$ layer inside the droplet at $\chi_{AB}=0$ ($z$ is the axis perpendicular to the surface, surface is located at $z=0$). The droplet was identified based on the threshold of density field; if $\phi_A(\textbf{r})$ or $\phi_B(\textbf{r})$ exceeded $0.01$ in a lattice cell, such cell was counted as within the droplet and the values of $\phi_A(\textbf{r})$ and $\phi_B(\textbf{r})$ were stored. $\phi_A^{surf}$ was equal to the ratio of the average density of $A$ segments per lattice cell within a droplet to the sum of average densities of $A$ and $B$ segments per lattice cell within a droplet. As a result, $Pref_A$ characterized the excess of $A$ monomer density compared to the diblock copolymer composition near the surface in a structureless droplet. $Pref_A\approx 0$ corresponds to the surfaces that do not prefer either block (average density of $A$ near the surface is approximately equal to just the composition of the diblock). $Pref_A>0$ and $Pref_A<0$ correspond to the surfaces that prefer $A$ block and $B$ block, respectively. Finally, $Pref_A$ was found to be approximately proportional to the difference between interaction parameters between the surface and the blocks: $Pref_A\propto \chi_{BS}N-\chi_{AS}N$ (Fig. \ref{fig_s_prefa}). This fact will be used later to set the values of $\chi_{ij}N$ necessary to reach a target $\theta$ and $Pref_A$.

Second, let us describe the procedure for the measurement of the contact angle of a BCP droplets at $\chi_{AB}=0$ (this quantity is denoted as $\theta$ in this work).
We developed a geometry-based approach to measure contact angles in 2D space SCFT calculations based on the spatial distribution of density $\phi_i(\textbf{r})$. After equilibration of the droplet at $\chi_{AB}=0$, we extracted $\phi_A(\textbf{r})$ and $\phi_B(\textbf{r})$ fields and identified the droplet boundary based on a threshold criterion. For low and intermediate $\theta$ (i.e., $\theta\lesssim92\degree$), the droplets could be approximated well by a sector of a circle. As a result, for these $\theta$, we fit a circle into the identified droplet boundary and obtained $\theta$ geometrically from the circle-substrate intersection points using inverse trigonometric functions. When a droplet wetted substrate weakly (high $\theta$), its shape became significantly acircular. Therefore, we fitted the circle just to the part of the droplet close to the surface to extract $\theta$. This approach provides robust measurements without sensitivity to local density fluctuations.

We mapped the six independent $\chi_{ij}N$ parameters onto four effective parameters $\chi_{AB}N$, $\chi_{AC}N$, $\theta$, and $Pref_A$ using Eq. \ref{eq:young}. The primary goal of this mapping was to make the four effective quantities independent: each of them should depend very weakly on the other effective quantities. By constructing this mapping, we were able to build a true four-dimensional morphological diagram that comprehensively explores the landscape of BCP nanodroplet morphologies. By varying the dependencies of $\chi_{AS}N$, $\chi_{BS}N$, and $\chi_{CS}N$ on $\chi_{AC}N$, we were able to control $\theta$ and $Pref_A$ in a wide range and independently of each other (Eq. \ref{eq:young} shows very small standard deviations of each quantity measured across all systems with other effective quantities varied).

The expressions in Eq. \ref{eq:young} were obtained by empirically correcting formulas obtained from the simplest Young's equation: $\gamma_{C,S}=\gamma_{S,BCP}+\gamma_{C,BCP}\cos\theta$. Since the dependence of interfacial tensions $\gamma_{i,j}$ between the system components $i$ and $j$ on $\chi_{ij}N$ is complex but direct \cite{hur2012chebyshev}, we used the simplest possible linear approximation as the initial guess before SCFT-informed correction: $\gamma_{C,S}=\chi_{CS}N$, $\gamma_{S,BCP}=(\chi_{AS}N+\chi_{BS}N)/2$, and $\gamma_{C,BCP}=(\chi_{AC}N+\chi_{BC}N)/2$. Since $Pref_A\approx 0.005(\chi_{BS}N-\chi_{AS}N)$ (see Fig. \ref{fig_s_prefa}), and $\chi_{BC}N=18$, we can rewrite the initial guess equation as follows: $\chi_{CS}N=\chi_{AS}N+Pref_A/0.0025+(\chi_{AC}N/2+9)\cos\theta$. For $Pref_A\neq 0$, we set $\chi_{AS}N$ and $\chi_{BS}N$ to a predetermined value to set $Pref_A$ and modified the coefficients in the $\chi_{CS}N(\chi_{AC}N)$ dependency for each $\theta$ and $Pref_A$ to set $\theta$ to a predetermined target value (independent of $Pref_A$ and $\chi_{AC}N$). For $Pref_A\approx0$ and $\theta\lesssim 92\degree$, we set $\chi_{AS}N=\chi_{BS}N=0$ and did the same procedure of modification. For $Pref_A\approx0$ and $\theta\approx 141\degree$, we set $\chi_{CS}N=0$ and modified the dependency of $\chi_{AS}N=\chi_{BS}N$ on $\chi_{AC}N$ to set $\theta$ to a target value (independent of $\chi_{AC}N$).

\small

\begin{equation}
\label{eq:young}
\begin{cases}
    Pref_A\approx -0.008\pm0.010,\ \theta\approx 67.4\pm1.8\degree: \chi_{CS}N=7\chi_{AC}N/18+5;\ \chi_{AS}=\chi_{BS}=0\\
    Pref_A\approx -0.008\pm0.010,\ \theta\approx 92.3\degree\pm1.5\degree: \chi_{CS}=\chi_{AS}=\chi_{BS}=0\\
    Pref_A\approx -0.008\pm0.010,\ \theta\approx 140.6\pm1.5\degree: \chi_{CS}=0;\ \chi_{AS}N=\chi_{BS}N=\chi_{AC}N/2+7\\
    Pref_A\approx 0.048\pm0.008,\ \theta\approx 67.4\pm1.8\degree: \chi_{CS}N=7\chi_{AC}N/18+9.5;\ \chi_{AS}=0;\ \chi_{BS}=9\\
    Pref_A\approx 0.048\pm0.008,\
    \theta\approx 92.3\degree\pm1.5\degree: \chi_{CS}N=4.5;\ \chi_{AS}=0;\ \chi_{BS}=9\\
    Pref_A\approx 0.048\pm0.008,\ \theta\approx 140.6\pm1.5\degree: \chi_{CS}N=9-\chi_{AC}N/2;\ \chi_{AS}=13.5;\ \chi_{BS}=22.5\\
    Pref_A\approx 0.094\pm0.007,\ \theta\approx 67.4\pm1.8\degree: \chi_{CS}N=7\chi_{AC}N/18+14;\ \chi_{AS}=0;\ \chi_{BS}=18\\
    Pref_A\approx 0.094\pm0.007,\ \theta\approx 92.3\degree\pm1.5\degree: \chi_{CS}N=9;\ \chi_{AS}=0;\ \chi_{BS}=18\\
    Pref_A\approx 0.094\pm0.007,\ \theta\approx 140.6\pm1.5\degree: \chi_{CS}N=11.5-\chi_{AC}N/2;\ \chi_{AS}=13.5;\ \chi_{BS}=31.5\\
    Pref_A\approx 0.157\pm0.006,\ \theta\approx 67.4\pm1.8\degree: \chi_{CS}N=7\chi_{AC}N/18+17.5;\ \chi_{AS}=0;\ \chi_{BS}=35\\
    Pref_A\approx 0.157\pm0.006,\ \theta\approx 92.3\degree\pm1.5\degree: \chi_{CS}N=12.5;\ \chi_{AS}=0;\ \chi_{BS}=35\\
    Pref_A\approx 0.157\pm0.006,\ \theta\approx 140.6\pm1.5\degree: \chi_{CS}N=15-\chi_{AC}N/2;\ \chi_{AS}=13.5;\ \chi_{BS}=48.5\\
    Pref_A\approx -0.064\pm0.013,\ \theta\approx 67.4\pm1.8\degree: \chi_{CS}N=7\chi_{AC}N/18+9.5;\ \chi_{AS}=9;\ \chi_{BS}=0\\
    Pref_A\approx -0.064\pm0.013,\ \theta\approx 92.3\degree\pm1.5\degree: \chi_{CS}N=4.5;\ \chi_{AS}=9;\ \chi_{BS}=0\\
    Pref_A\approx -0.064\pm0.013,\ \theta\approx 140.6\pm1.5\degree: \chi_{CS}N=9-\chi_{AC}N/2;\ \chi_{AS}=22.5;\ \chi_{BS}=13.5\\
    \end{cases}
\end{equation}

\normalsize

\section{3D Space SCFT Calculations Method}

In this section, we will outline the procedure of the SCFT calculations in 3D space (the results are presented in section 2.2 of the main text). As a first step, we created analogous 3D structures from the selected equilibrium 2D droplet morphologies (Fig. 2 in the main text) as follows. The simulation cell was set to have dimensions $x\times y\times z$: $20R_g\times20R_g\times8.4R_g$; grid cell size was set similarly to the 2D space calculations: $0.1R_g\times 0.1R_g\times 0.1R_g$. First, we copied the chemical potential and density fields of a 2D structure along the newly created $y$ axis from $y=10R_g-y_d/2$ to $y=10R_g+y_d/2$. The space between $y=0$ and $y=10R_g-y_d/2$ as well as between $y=10R_g+y_d/2$ and $y=20R_g$ was filled with pure $C$ homopolymer. $y_d=6.4R_g$ was set such that the volume of the 3D droplet at $\chi_{AB}=0$ and $\theta\approx92\degree$ conditions was approximately equal to the volume of a hemisphere with the radius equal to the radius of the 2D droplet at the same conditions: $V_{droplet}\approx 192 R_g^3$. To achieve this $V_{droplet}$, we set the BCP volume fraction to $0.057143$. This choice of the 3D droplet volume ensured that the confinement conditions in 3D droplets are analogous to those in 2D droplets. The volume fraction of $S$ homopolymer was left at $0.04$ to mimic the same surface as in 2D space calculations. Second, after preparing this artificial state, we used it as initial condition for the real-space SCFT calculations that produced stabilized nanodroplets in 3D space analogous to those obtained in 2D space (Fig. 4 in the main text). The SCFT equations solution algorithm and free energy stabilization criteria were the same as for 2D space calculations (see section 4.1 in the main text).

After that, we selected a 3D structure and, at these interaction parameters, performed 3D space SCFT calculations initialized from all other structures in the diagram. This "brute force" algorithm ensured that the lowest free energy structure obtained after such calculations corresponded to the true equilibrium at the selected interaction parameters (if the initially obtained 3D structures represent all possible equilibrium single-droplet 3D structures in the studied range of interaction parameters).

The latter statement was supported for the analogous 2D droplets: the 2D structures used as templates for the creation of the 3D structures should comprise the entire set of equilibrium 2D structures in the studied interaction parameter range (see section 4.2 in the main text). As a result, we assumed that the created 3D structures should also represent the entire variety of equilibrium 3D structures in the studied range due to similar confinement conditions ensured by the choice of $V_{droplet}$. To confirm this further, we took three metastable 3D morphologies obtained after the brute-force equilibration algorithm and used them as initial structures for SCFT calculations at all studied interaction parameters. These structures remained metastable at all studied parameters. Such additional check ensured that the artificially created 3D structures likely comprised the entire set of equilibrium morphologies in the studied range of interaction parameters. As a result, the diagram obtained after the brute-force equilibration algorithm should be in equilibrium (Fig. 4 in the main text).

\section{Definition of a BCP Nanodroplet Morphology}

In order to perform the iterative correction algorithm as well as to analyze the structural complexity of nanodroplets, one has to evaluate at which values of interaction parameters one droplet morphology switches into another. We assumed that these transitions are of the first-order type, which means that one can imagine them as a process of competition between different local free energy minima corresponding to different morphologies. In 2D space SCFT diagrams, at a fixed $Pref_A$, we varied $\chi_{AB}N$ and $\chi_{AC}N$ values with $\Delta\chi_{AB}N=\Delta\chi_{AC}N=1$ step, and we set $\theta$ values to one of the three values: $\theta=\{67\degree,92\degree,141\degree\}$. Unfortunately, these steps with which the control parameters were changed were too large to detect the transition from one droplet morphology to another by monitoring a jump in free energy or its derivative.

As a result, we adopted a different definition of the morphological transition based on the ability of structures to transition into each other upon small changes in control parameters. First, one obtains a morphological diagram after performing real-space SCFT calculations in the given range of parameters $\chi_{AC}N$, $\chi_{AB}N$, and $\theta$. We will further call the structures obtained in this diagram as "original" morphologies. Then, one can take the original structure obtained at $\{\chi_{AC}^{(1)}N,\chi_{AB}^{(1)}N,\theta^{(1)}\}$ parameters (we will call this morphology as "morphology 1") and use it as an initial structure for the real-space SCFT calculations at any of the following six parameter triplets $\{\chi_{AC}^{(1)}N\pm1,\chi_{AB}^{(1)}N\pm1,\theta^{(1)}\}$, $\{\chi_{AC}^{(1)}N,\chi_{AB}^{(1)}N,\theta\neq\theta^{(1)}\}$. We assume that the step sizes $\Delta\chi_{AC}N$, $\Delta\chi_{AB}N$, $\Delta\theta$ are small enough, i.e., such calculation should begin in the vicinity of a local minimum corresponding to the morphology 1 at each of the six parameter triplets. Therefore, if at one of these triplets the newly calculated structure does not coincide with the original morphology, the original morphology lies in a different free energy minimum than morphology 1. As a result, there exists a transition between morphology 1 and the original morphology at this parameter triplet.

Theoretically, one can perform such calculations for each point in the diagram and determine at which triplets $\{\chi_{AC}^{(1)}N,\chi_{AB}^{(1)}N,\theta^{(1)}\}$ there exist morphological transitions. After this, the morphological regions can be defined as maximal connected regions without phase transitions in the $\{\chi_{AC}N,\chi_{AB}N,\theta\}$ parameter space. However, this method is computationally infeasible, since it requires six times more SCFT calculations than originally performed to build the diagram (i.e., in this work, more than $10^4$ calculations at just the first iteration of the iterative correction algorithm).

Thus, we designed a separate approximate methodology of dividing the diagram into regions. A simpler version of this method was applied to study transitions between BCP nanoparticles in solution in our previous work \cite{petrov2025symmetric}. After performing SCFT calculations in the studied parameter range, we subdivided the diagram of states into regions in which structures had similar distribution of density of each species. After this, we selected one or two pairs of structures lying at the transition surface between a pair of identified regions and performed the SCFT calculations that took one of the structures in the pair as an initial structure and where the interaction parameters are set as for the other structure in the pair. If these pairs of structures did not transform into each other via such calculations, we concluded that the transition points were identified correctly. Such a check was performed for all identified regions and transition surfaces in the diagram.

After this, we showed that most likely there exist no unidentified transition surfaces within each initially identified region. In principle, this would require the transformation of all structures belonging to the same region into each other via small changes in interaction parameters (as described in paragraph 2 of this section). Due to the computational infeasibility of such a direct test (as discussed above), we selected a region and identified several pairs of structures obtained at maximally different interaction parameters (i.e., lying in the opposite "corners" of the region). After this, we performed real-space SCFT calculations that took one of the structures in the pair as an initial structure and where the interaction parameters are set as for the other structure in the pair. If there existed a pair of structures that did not transform into each other via such calculations, we gave the pair "a second chance". In particular, we selected one structure from the pair and equilibrated it at $\chi_{AC}N$ value of the second structure in the pair. Then, we equilibrated the obtained structure at $\theta$ value of the second structure (if the difference between $\theta$ of the first and second structure exceeded $\Delta\theta$, the obtained structure was first equilibrated at $\theta\approx92\degree$ and then at the $\theta$ value of the second structure). Finally, the obtained structure was equilibrated at the interaction parameters of the second structure.

As a result, if all maximally distant pairs within the selected region were transformable into each other via either of the two methods described above, we assumed that all other structures in that region are transformable into each other as well (since all other structures were obtained at more similar interaction parameters than these distant pairs). Thus, we concluded that most likely there are no unidentified morphological transitions within this region. If neither of the transformation methods worked for all distant pairs of structures inside the studied region, we subdivided that region into smaller regions and performed the "transformability check" again. We performed the procedure described above for all regions in the diagram until we reached the diagram in which all structures inside each region were deemed transformable into each other and in which the transition points were identified correctly (as described above).

\section{Equilibrium Morphological Regions of Non-Adsorbed Droplets}

At $Pref_A\approx0$ and $\theta\approx 141\degree$, some of the structures desorbed from the surface at equilibrium (Fig. 2a in the main text). To identify the equilibrium morphological regions for each of the desorbed structures, we performed the following calculations. At $Pref_A\approx0$, $\theta\approx141\degree$, $\chi_{AB}N\in[10,50]$, and $\chi_{AC}N\in[0,9]$ we performed 2D space SCFT calculations of the desorbed nanodisk, desorbed onion-like particle, and desorbed two-micelle state. These states were found to be the only equilibrium nanoparticle morphologies in solution at these interaction parameters for this droplet size \cite{petrov2025symmetric}. $\theta\approx141\degree$ and $\chi_{AC}N\in[0,9]$ were chosen since these wetting/surrounding medium interaction conditions are the most favorable for nanodroplet desorption from the surface. SCFT calculations were performed in a larger simulation cell having dimensions $20R_g\times20R_g$ to eliminate the effect of the "ceiling" of the simulation cell, which is strong for the desorbed droplet states in the standard simulation cell $20R_g\times 8.4R_g$ used in all other calculations.

We compared the free energies of the aforementioned desorbed morphologies to the free energies of four surface-adsorbed structures that survived after the iterative correction algorithm (ICA) execution in that region of interaction parameters. These structures included the "standing disk", "onion", "$3A2B$ striped ellipsoid" structures (see Fig. 2a in the main text) as well as another metastable structure resembling the "$3A2B$ striped ellipsoid" structure but with one B-stripe larger than the other one. After performing SCFT calculations in the large cell of all aforementioned seven structures in the range of parameters mentioned above, we selected the lowest free energy structure as an equilibrium one. At $\chi_{AC}N\geq 5$, all equilibrium structures were surface-adsorbed. The boundaries between the "standing disk" and "onion" morphologies at $\chi_{AC}N=8,9$ coincided within $\Delta\chi_{AB}N=1$ error with the ones established by the ICA in the standard simulation cell, confirming the weakness of the "ceiling" effect for all adsorbed nanodroplets studied in this work in the standard simulation cell.

In addition, the execution of the same algorithm at $Pref_A\approx0$, $\theta\approx92\degree$, $\chi_{AB}N\in[10,50]$, $\chi_{AC}=0$ led to the absence of equilibrium non-adsorbed droplet morphologies. As a result, such structures are guaranteed to not exist in equilibrium at $\theta\lesssim90\degree$ (Fig. 2b,c in the main text).

\section{Diagrams Obtained During Iterative Correction Algorithm}

\begin{figure}[H]
\centering
  \begin{subfigure}{0.49\textwidth}
	\includegraphics[width=\linewidth,height=\textheight,keepaspectratio]{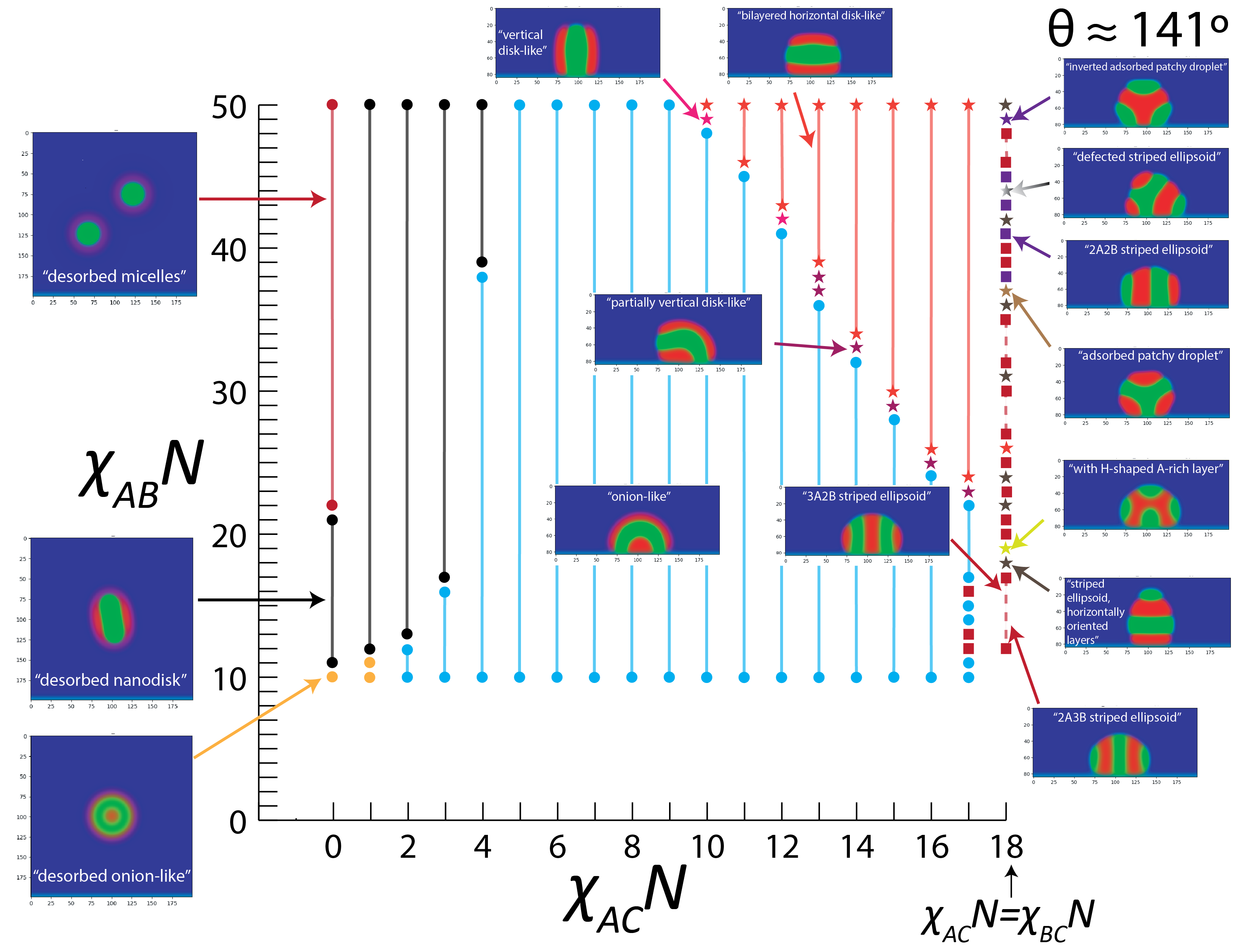}
	\caption{}
	\end{subfigure}
	\begin{subfigure}{0.49\textwidth}
	\includegraphics[width=\linewidth,height=\textheight,keepaspectratio]{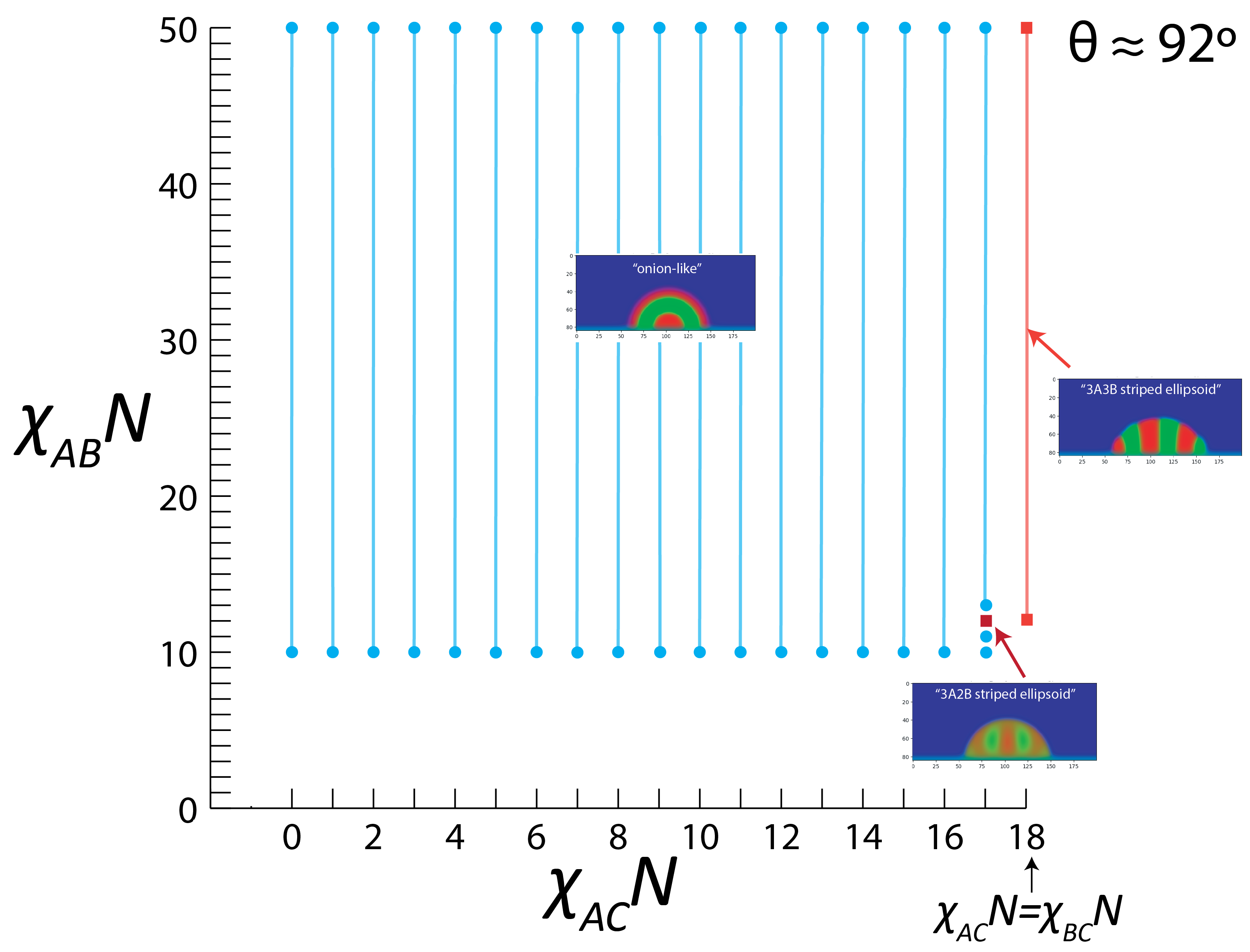}
	\caption{}
	\end{subfigure}
	\begin{subfigure}{0.49\textwidth}
	\includegraphics[width=\linewidth,height=\textheight,keepaspectratio]{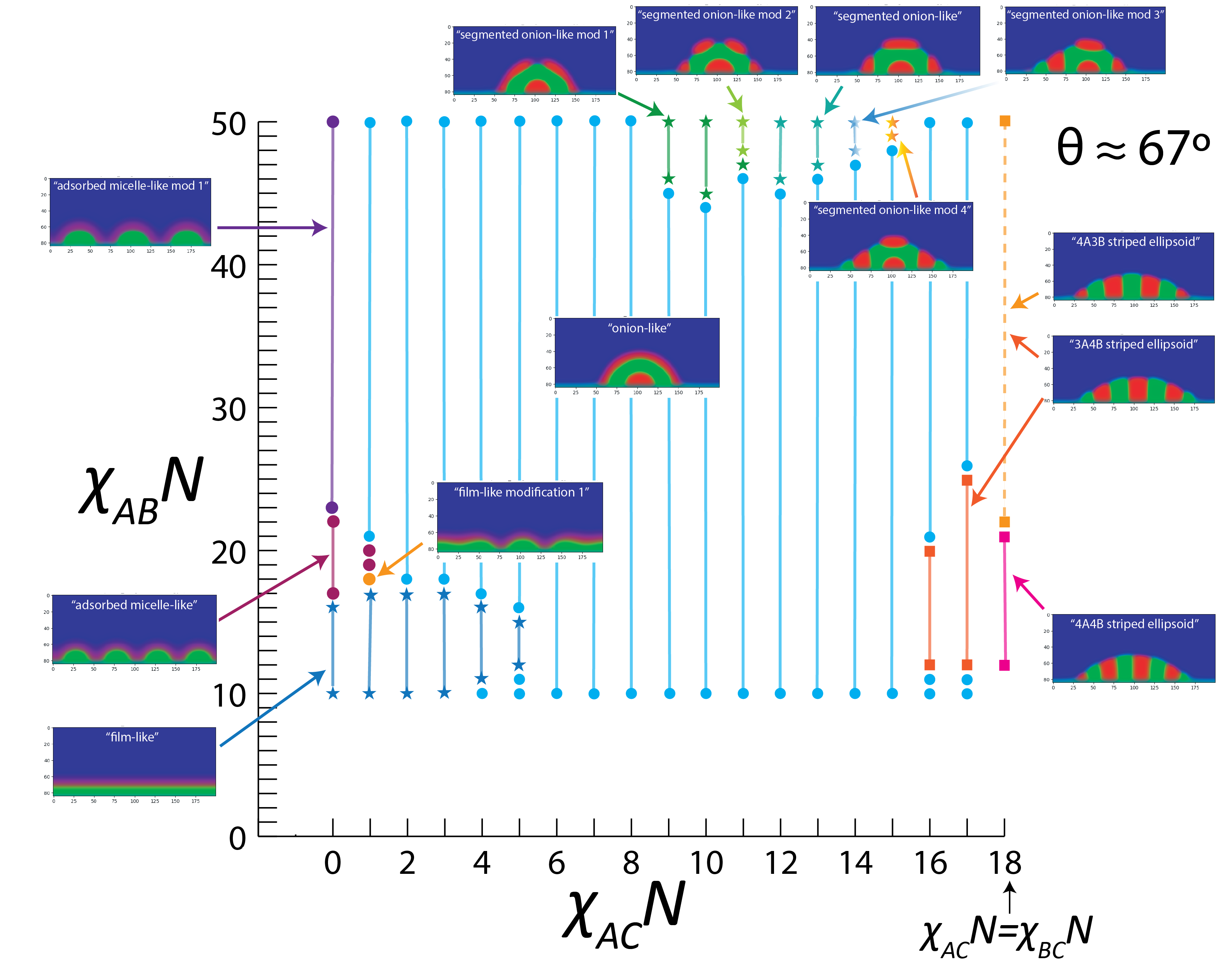}
	\caption{}
	\end{subfigure}
  \caption{Initial morphological diagram of nanodroplets adsorbed at a surface with no preference towards any of the blocks ($Pref_A\approx 0$). This diagram was used as an input for iterative correction algorithm. SCFT calculations were performed in 2D space. Diagrams (a), (b), and (c) are obtained at $\theta\approx 141\degree$, $\theta\approx 92\degree$, and $\theta\approx 67\degree$, respectively. Points of different colors represent the boundaries of different morphologies at a fixed $\chi_{AC}N$. Continuous lines between the boundary points show that this morphology is obtained at all values of $\chi_{AB}N$ between these points. A characteristic snapshot for each structure as well as its short name are shown for each morphological region. $A$ blocks are shown in red, $B$ blocks are colored green, the surface-tethered $S$ homopolymers are shown in light blue, and homopolymers $C$ modeling the surrounding medium are colored dark blue. Dashed lines represent the alternating presence of two morphologies between the boundary points at the given $\chi_{AC}N$. Since these morphologies did not differ in the number of lamellar layers but only in their order (for example, "$3A2B$" and "$2A3B$" striped ellipsoidal particle in (a)) and since $\chi_{AC}=\chi_{BC}$ along the dashed lines, there was no preference towards any of the two morphologies and they alternated randomly at these conditions.}
\label{fig_s_0pref_iter0}
\end{figure}

\begin{figure}[H]
\centering
  \begin{subfigure}{0.49\textwidth}
	\includegraphics[width=\linewidth,height=\textheight,keepaspectratio]{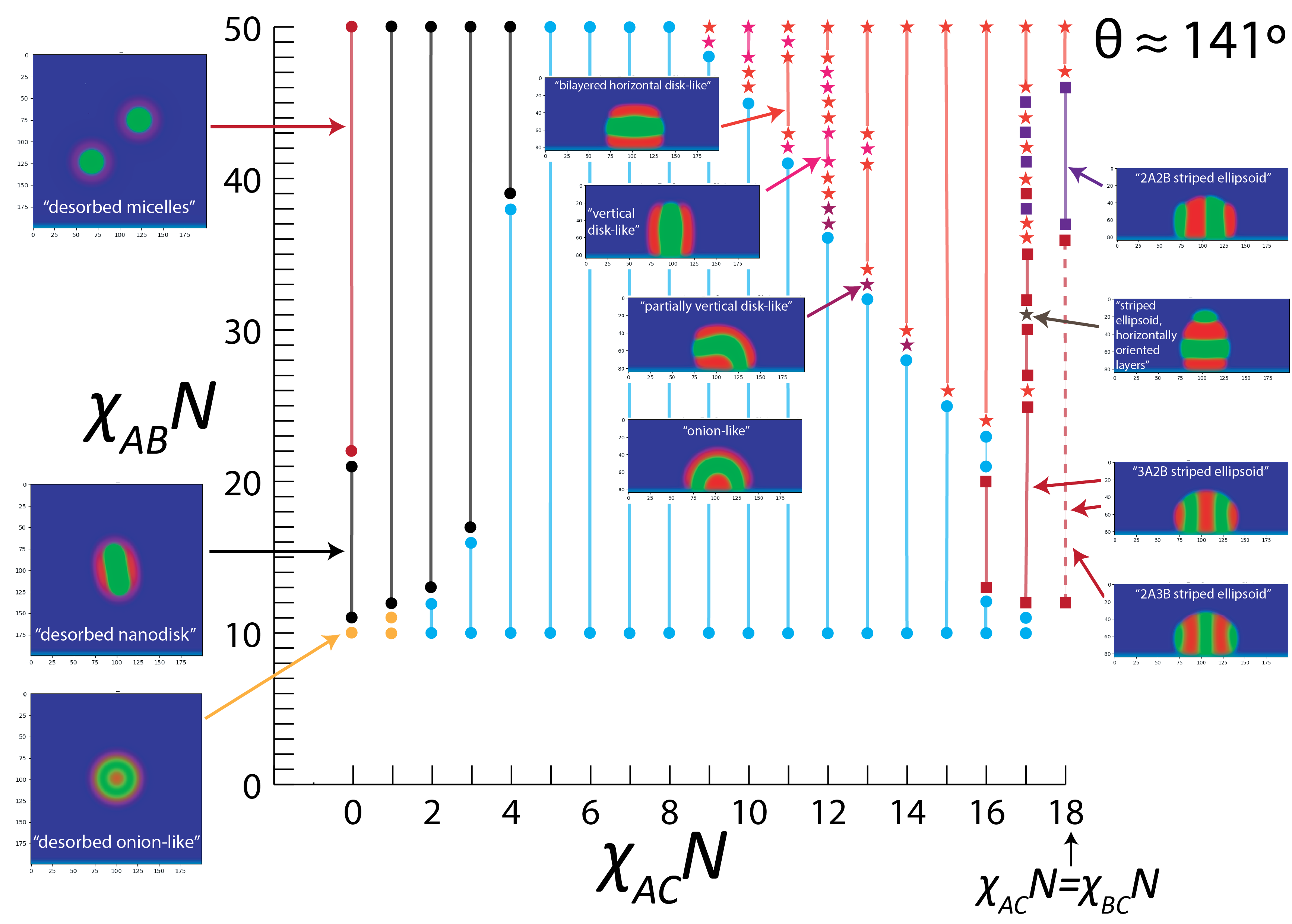}
	\caption{}
	\end{subfigure}
	\begin{subfigure}{0.49\textwidth}
	\includegraphics[width=\linewidth,height=\textheight,keepaspectratio]{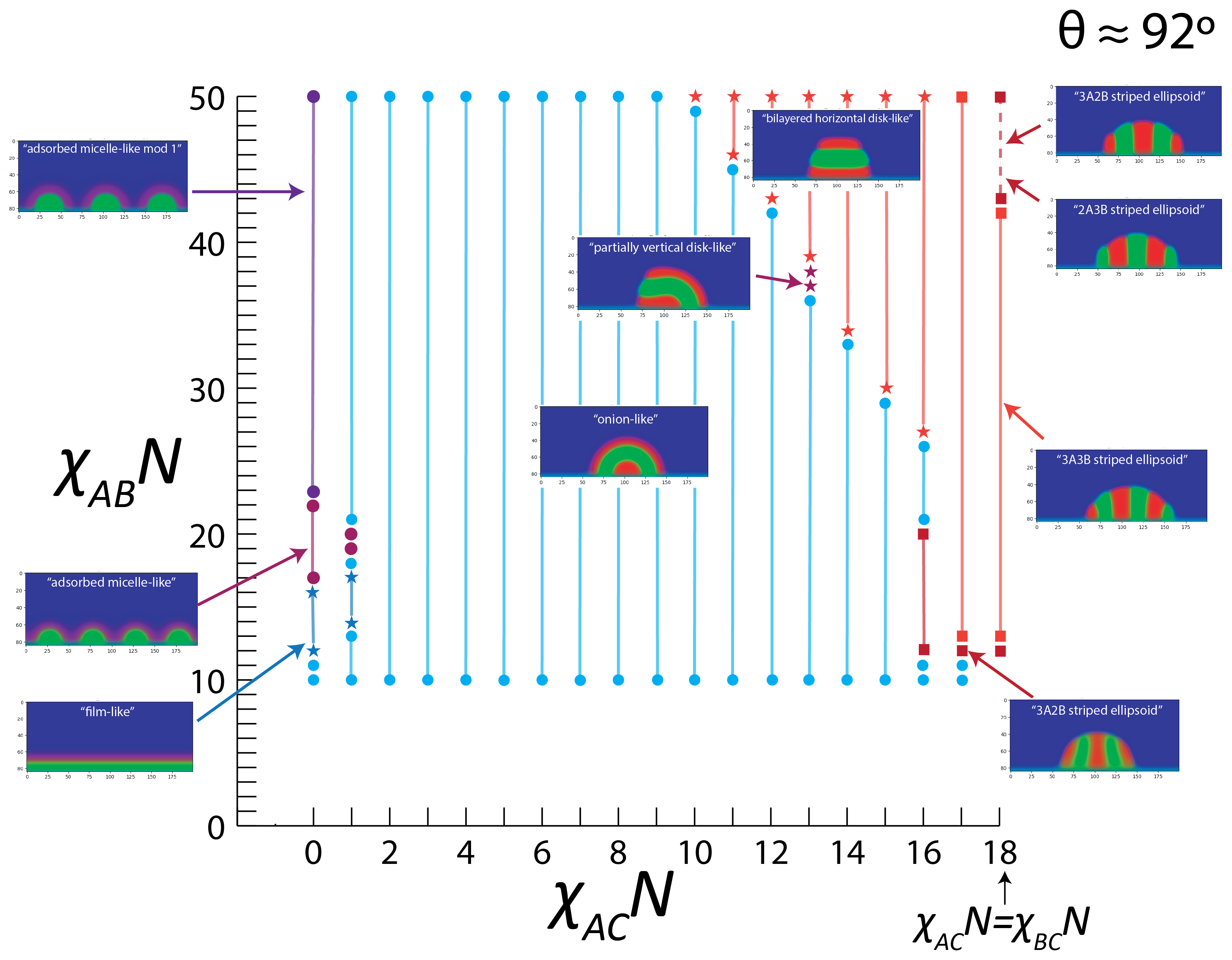}
	\caption{}
	\end{subfigure}
	\begin{subfigure}{0.49\textwidth}
	\includegraphics[width=\linewidth,height=\textheight,keepaspectratio]{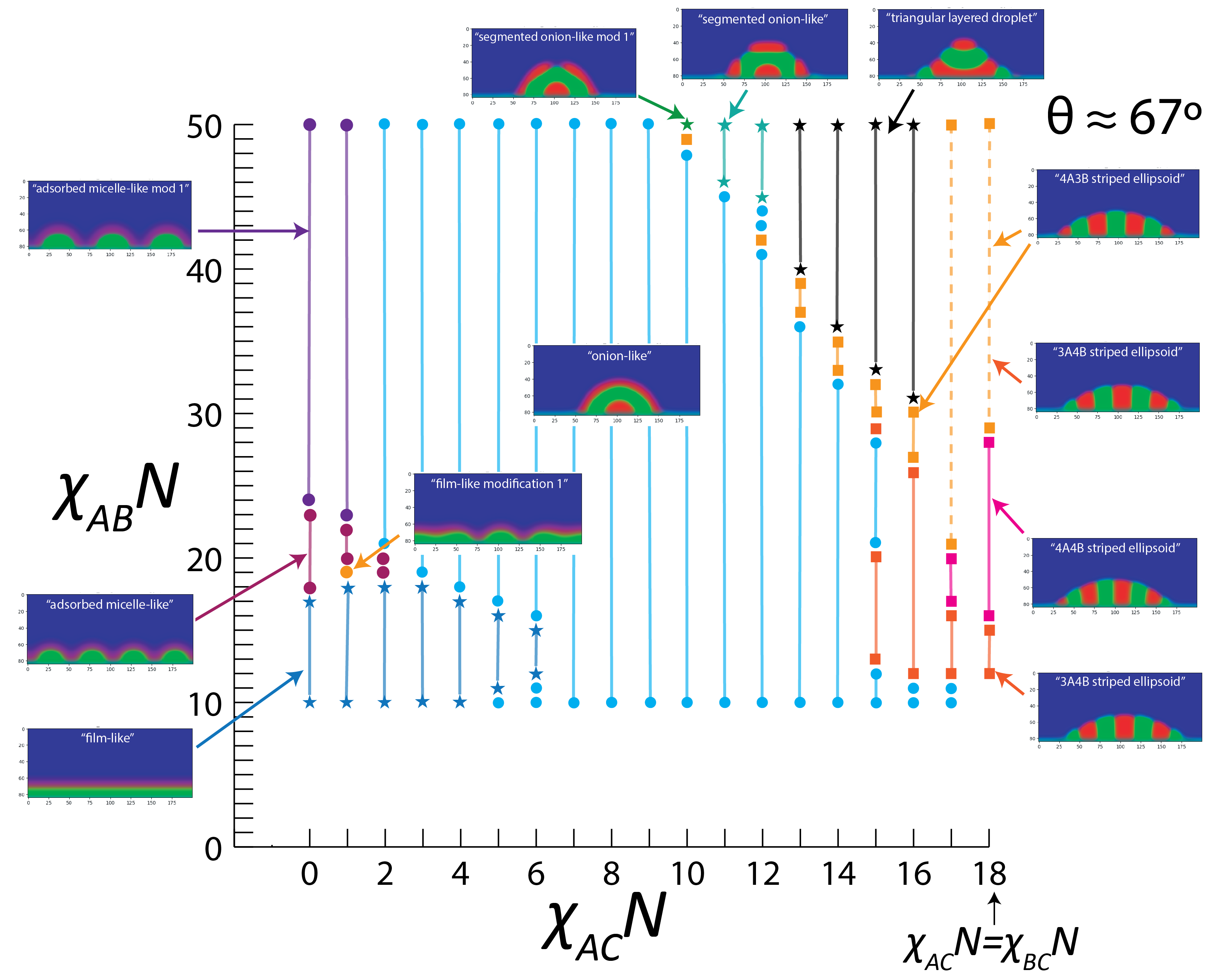}
	\caption{}
	\end{subfigure}
  \caption{Morphological diagram of nanodroplets adsorbed at a surface with no preference towards any of the blocks ($Pref_A\approx 0$); this diagram was obtained after the first iteration of the correction algorithm. SCFT calculations were performed in 2D space. Diagrams (a), (b), and (c) are obtained at $\theta\approx 141\degree$, $\theta\approx 92\degree$, and $\theta\approx 67\degree$, respectively. Points of different colors represent the boundaries of different morphologies at a fixed $\chi_{AC}N$. Continuous lines between the boundary points show that this morphology is obtained at all values of $\chi_{AB}N$ between these points. A characteristic snapshot for each structure as well as its short name are shown for each morphological region. $A$ blocks are shown in red, $B$ blocks are colored green, the surface-tethered $S$ homopolymers are shown in light blue, and homopolymers $C$ modeling the surrounding medium are colored dark blue. Dashed lines represent the alternating presence of two morphologies between the boundary points at the given $\chi_{AC}N$. These morphologies did not differ in the number of lamellar layers but only in their order (for example, "$3A2B$" and "$2A3B$" striped ellipsoidal particle in (a)).}
\label{fig_s_1pref_iter0}
\end{figure}

\section{Additional Plots}

\begin{figure}[H]
\centering
  \includegraphics[width=0.6\linewidth]{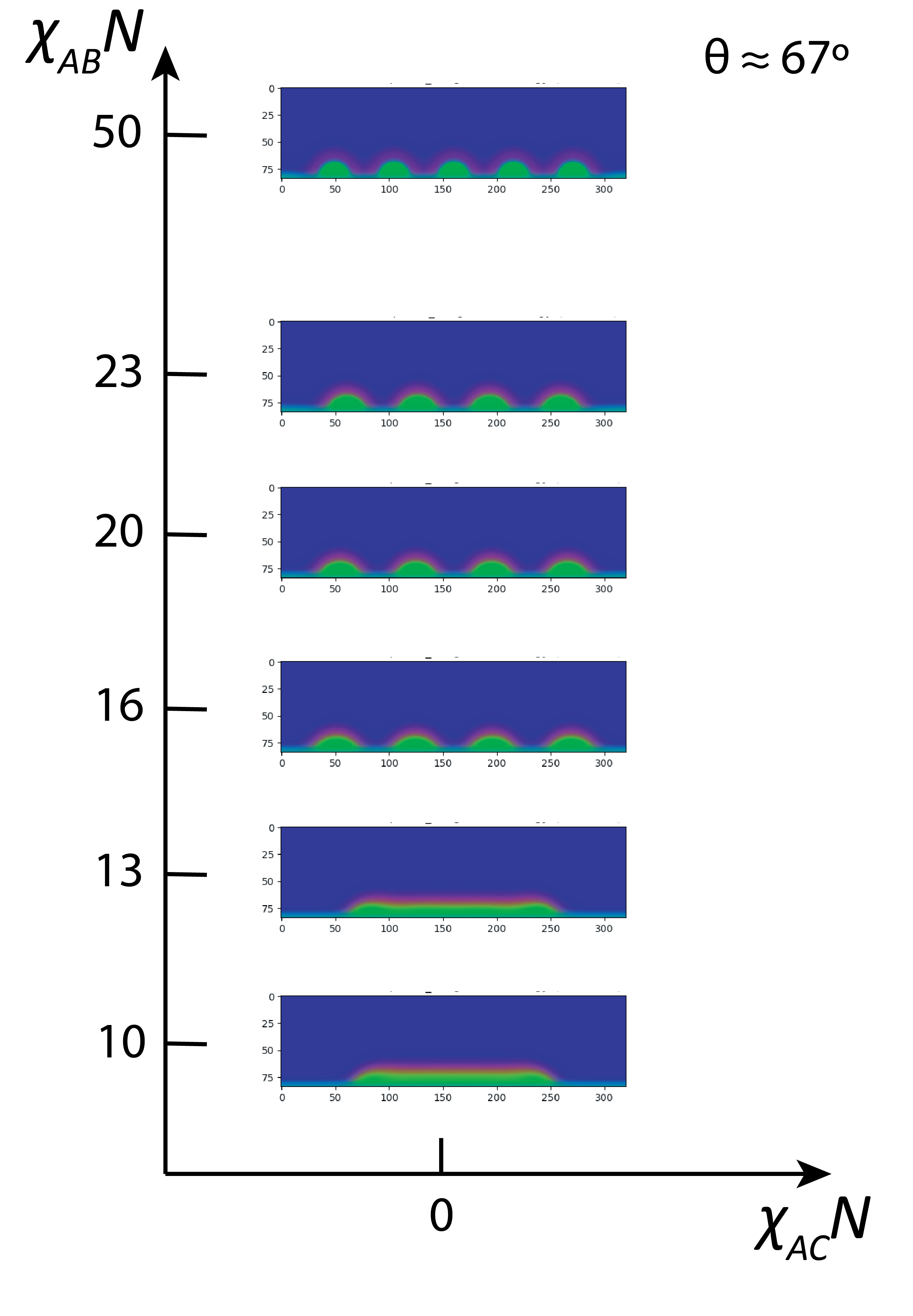}
  \caption{Near-equilibrium structures formed by symmetric diblock copolymers adsorbed on a strongly wetting neutral surface ($Pref_A\approx 0$, $\theta\approx67\degree$) at $\chi_{AC}=0$ (full compatibility of block $A$ with the surrounding medium). $A$ blocks are shown in red, $B$ blocks are colored green, the surface-tethered $S$ homopolymers are shown in light blue, and homopolymers $C$ modeling the surrounding medium are colored dark blue. The structures were obtained in a calculation cell with larger $x$-dimension than used in this work for all other systems ($32R_g\times 8.4 R_g$). The calculations proceeded as follows. First, a structureless surface-adsorbed droplet was obtained at $\chi_{AB}=\chi_{AC}=0$ conditions. Second, this structure was used as an initial structure for real-space SCFT calculations performed as described in Methods at different values of $\chi_{AB}N$. Third, the structures obtained after such calculations were used as initial structures for the real-space calculations at every studied value of $\chi_{AB}N$; the final near-equilibrium structure was selected as the lowest free energy solution.}
  \label{fig_s_extendedbox}
\end{figure}

\begin{figure}[H]
\centering
  \begin{subfigure}{0.49\textwidth}
	\includegraphics[width=\linewidth,height=\textheight,keepaspectratio]{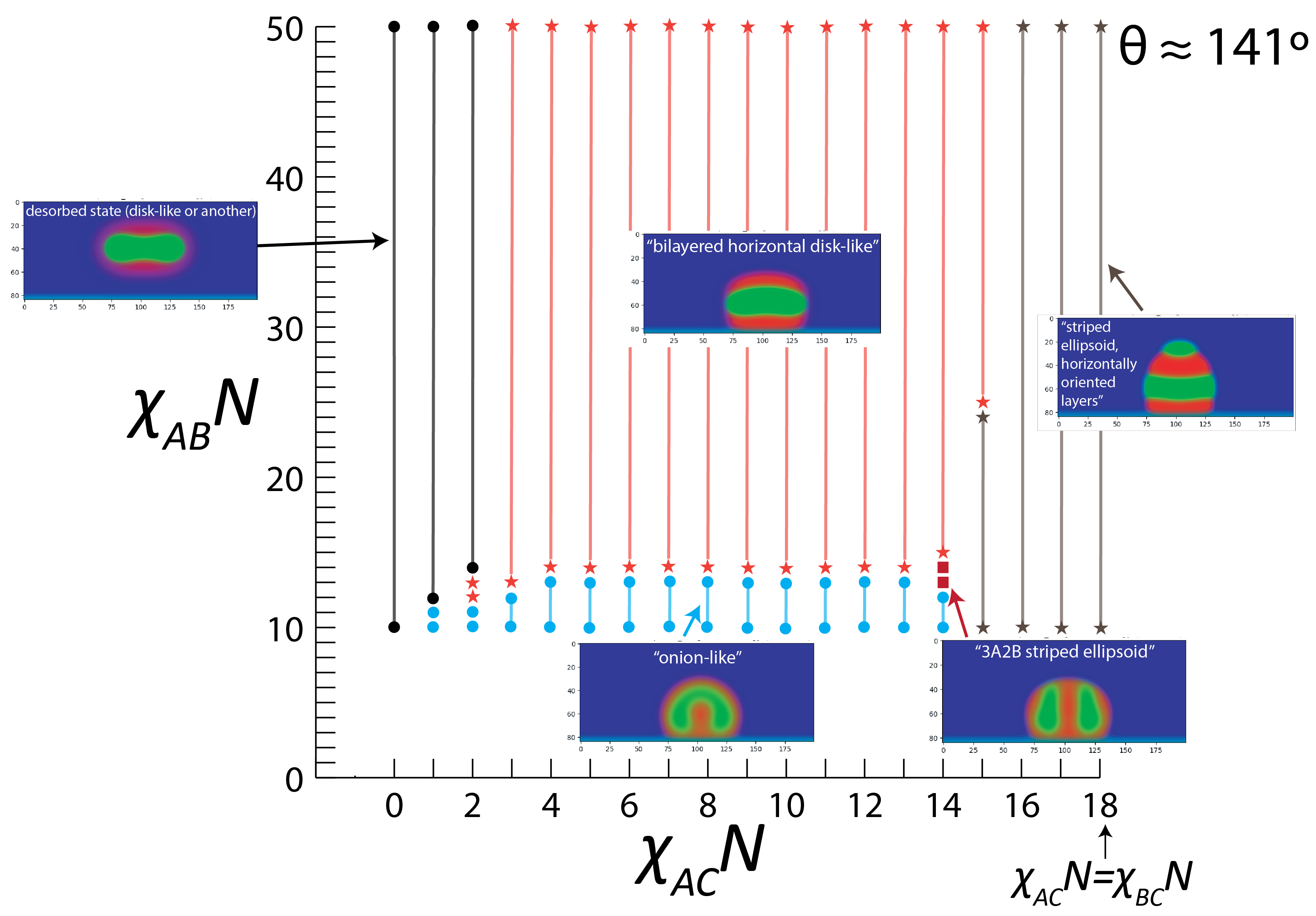}
	\caption{}
	\end{subfigure}
	\begin{subfigure}{0.49\textwidth}
	\includegraphics[width=\linewidth,height=\textheight,keepaspectratio]{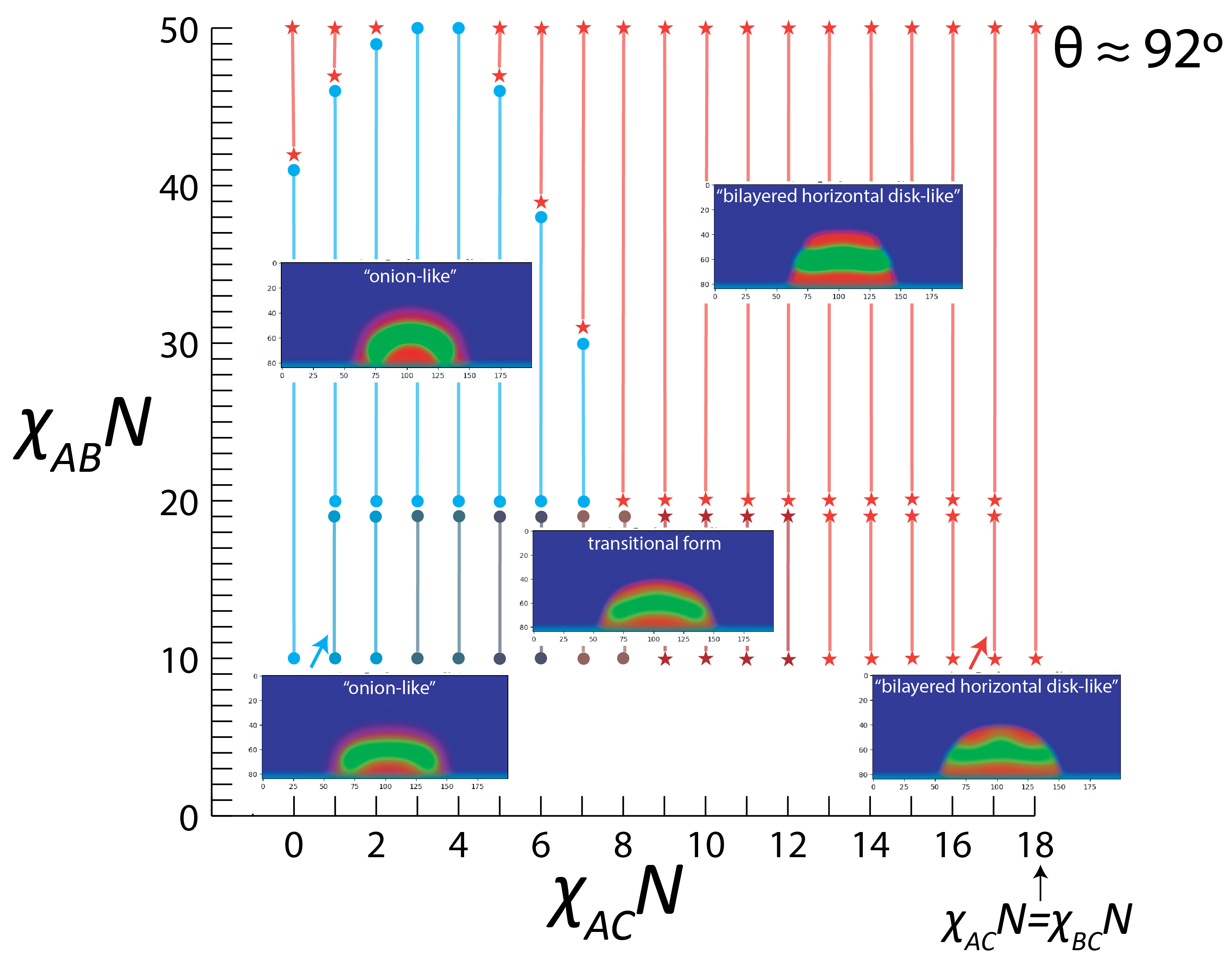}
	\caption{}
	\end{subfigure}
	\begin{subfigure}{0.49\textwidth}
	\includegraphics[width=\linewidth,height=\textheight,keepaspectratio]{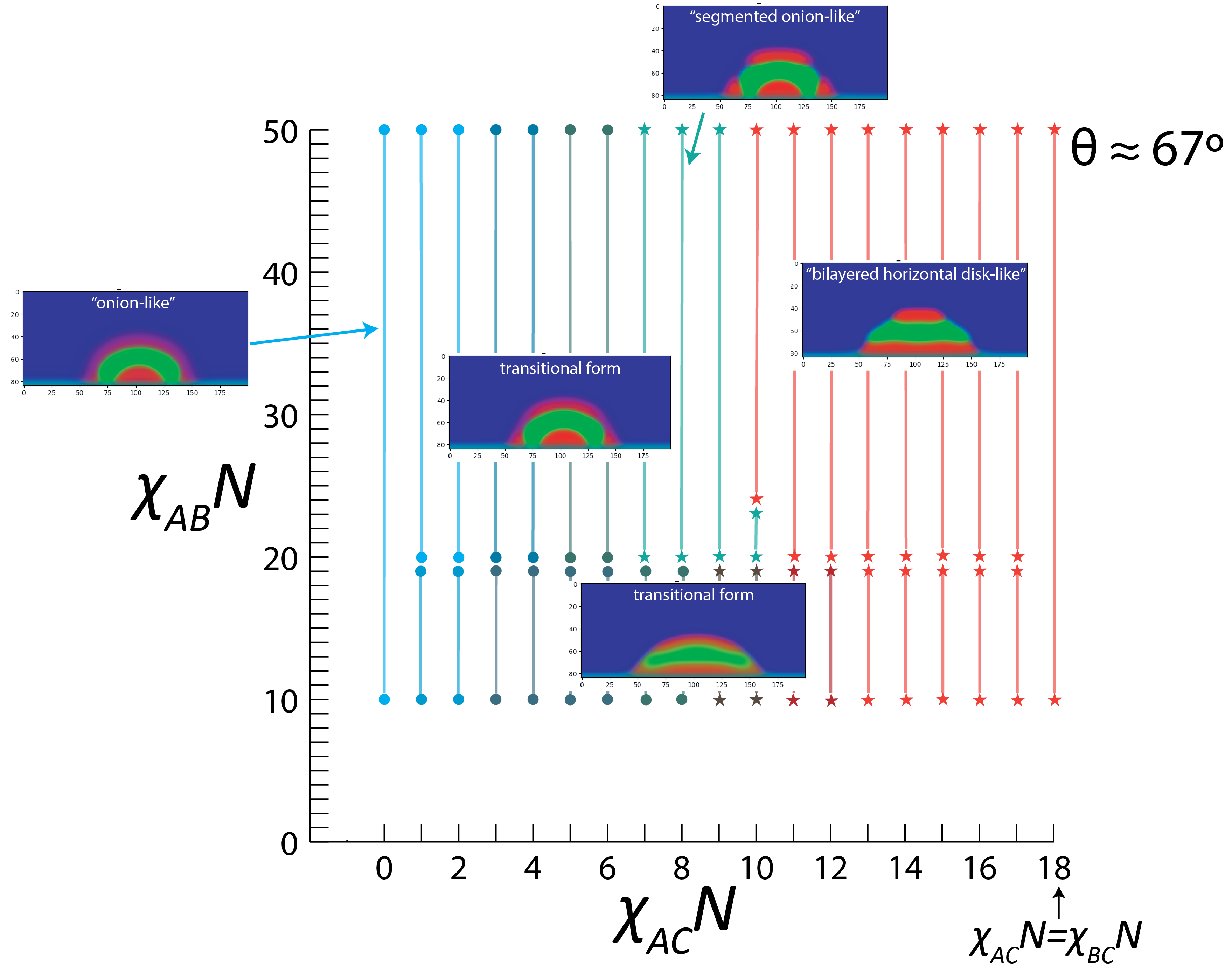}
	\caption{}
	\end{subfigure}
  \caption{Morphological diagram of nanodroplets adsorbed at a surface with "weak" preference towards $A$ blocks ($Pref_A\approx 0.048$); the surrounding medium is either neutral or preferential to the same block ($\chi_{AC}\leq\chi_{BC}$). SCFT calculations were performed in 2D space. Diagrams (a), (b), and (c) are obtained at $\theta\approx 141\degree$, $\theta\approx 92\degree$, and $\theta\approx 67\degree$, respectively. Points of different colors represent the boundaries of different morphologies at a fixed $\chi_{AC}N$. Continuous lines between the boundary points show that this morphology is obtained at all values of $\chi_{AB}N$ between these points. For $\theta\leq92\degree$ and $\chi_{AB}N\lesssim20$, the onion-like droplet transitioned into disk-like structure smoothly. Another smooth transitional region was found for $\theta\approx67\degree$, $\chi_{AC}N\lesssim7$, and $\chi_{AB}N\gtrsim20$ between the onion-like droplet and segmented onion-like state. We did not perform additional procedures to determine an exact boundary between these morphologies in these regions, since we did not carry out the iterative correction of this diagram. As a result, these transitional regions are shown by points with color changing by a gradient between the color corresponding to the onion-like droplet and the color corresponding to the disk-like structure or segmented onion-like state. A characteristic snapshot for each structure as well as its short name are shown for each morphological region. $A$ blocks are shown in red, $B$ blocks are colored green, the surface-tethered $S$ homopolymers are shown in light blue, and homopolymers $C$ modeling the surrounding medium are colored dark blue.}
\label{fig_s_weakpref}
\end{figure}

\begin{figure}[H]
\centering
  \begin{subfigure}{0.49\textwidth}
	\includegraphics[width=\linewidth,height=\textheight,keepaspectratio]{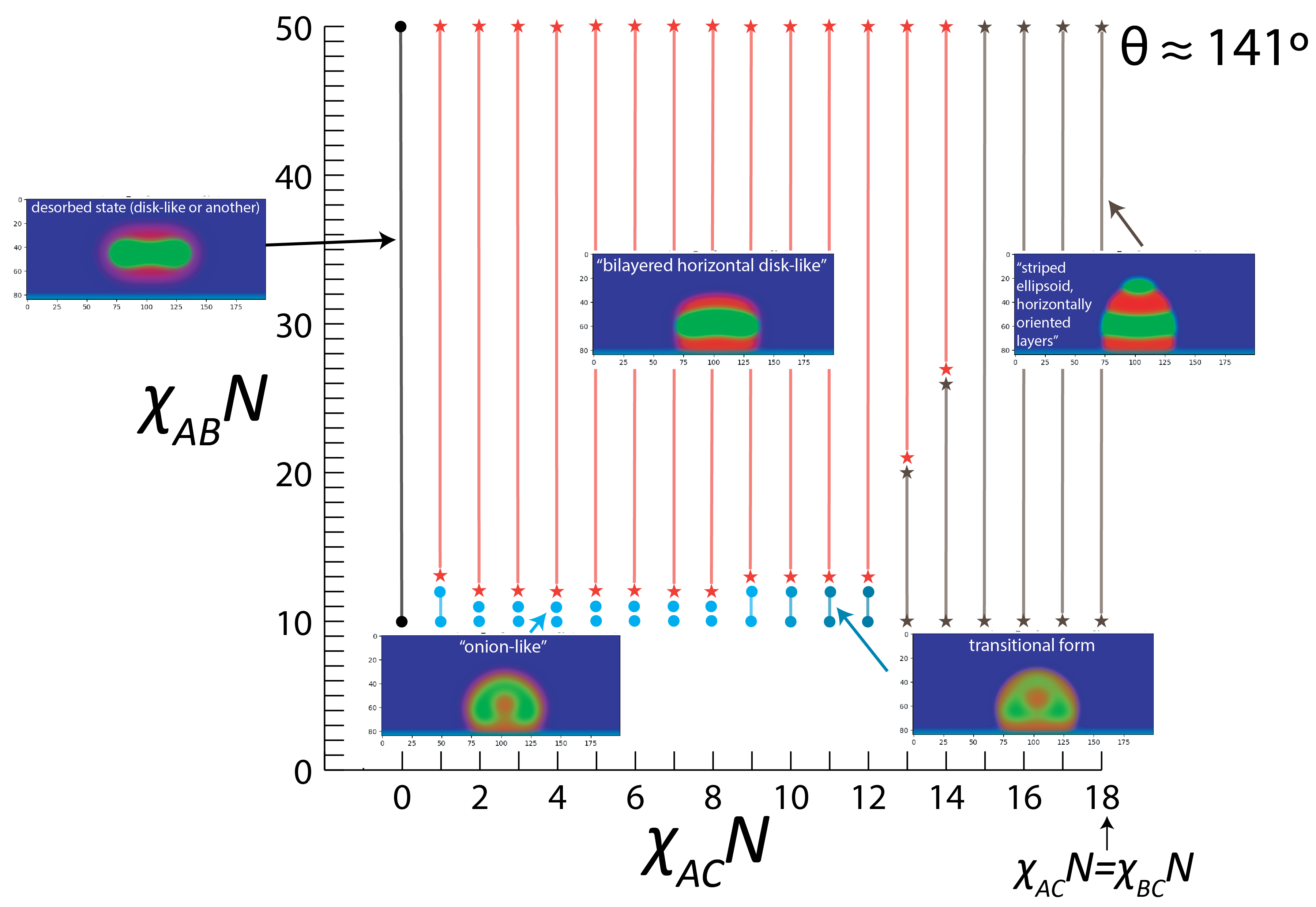}
	\caption{}
	\end{subfigure}
	\begin{subfigure}{0.49\textwidth}
	\includegraphics[width=\linewidth,height=\textheight,keepaspectratio]{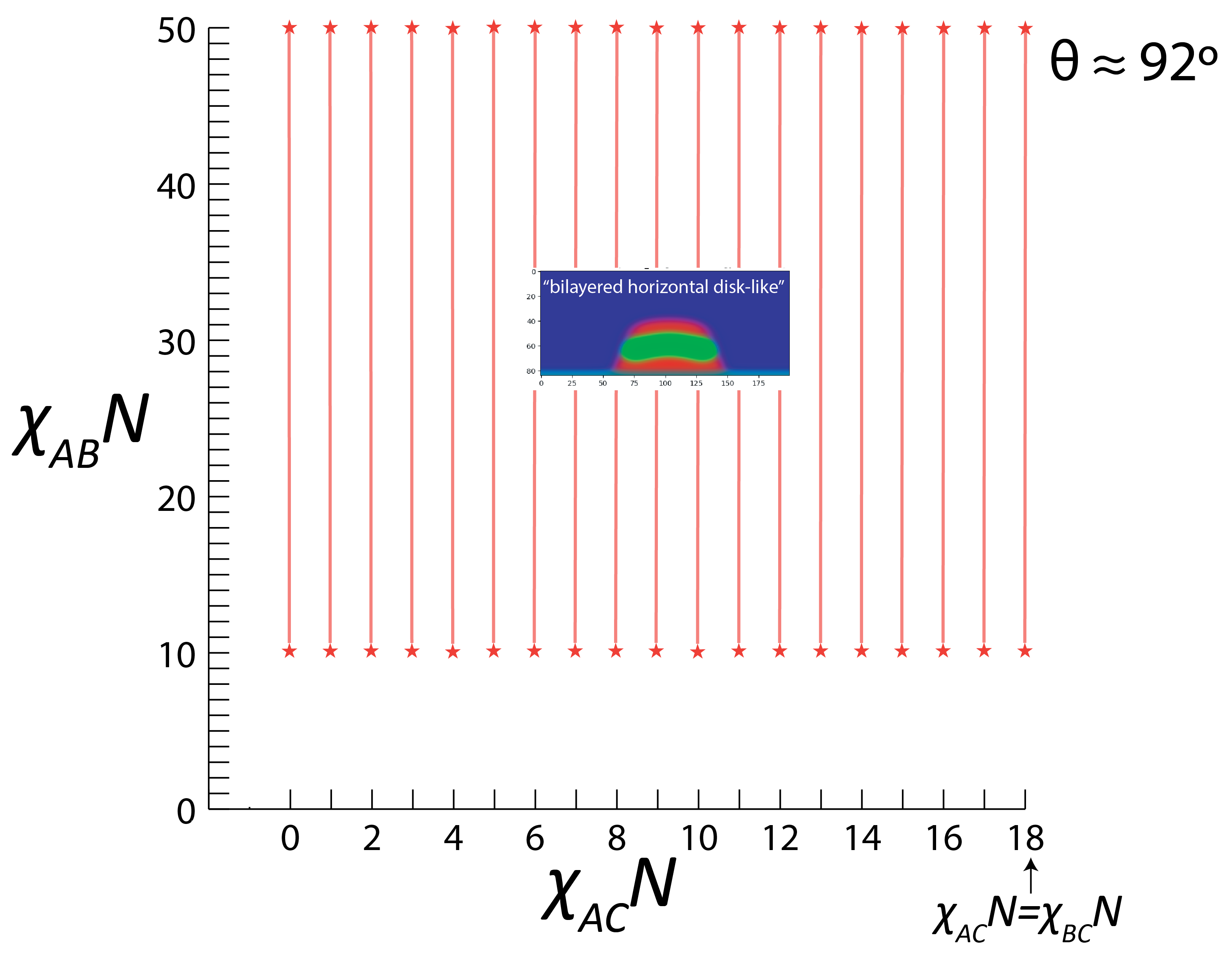}
	\caption{}
	\end{subfigure}
	\begin{subfigure}{0.49\textwidth}
	\includegraphics[width=\linewidth,height=\textheight,keepaspectratio]{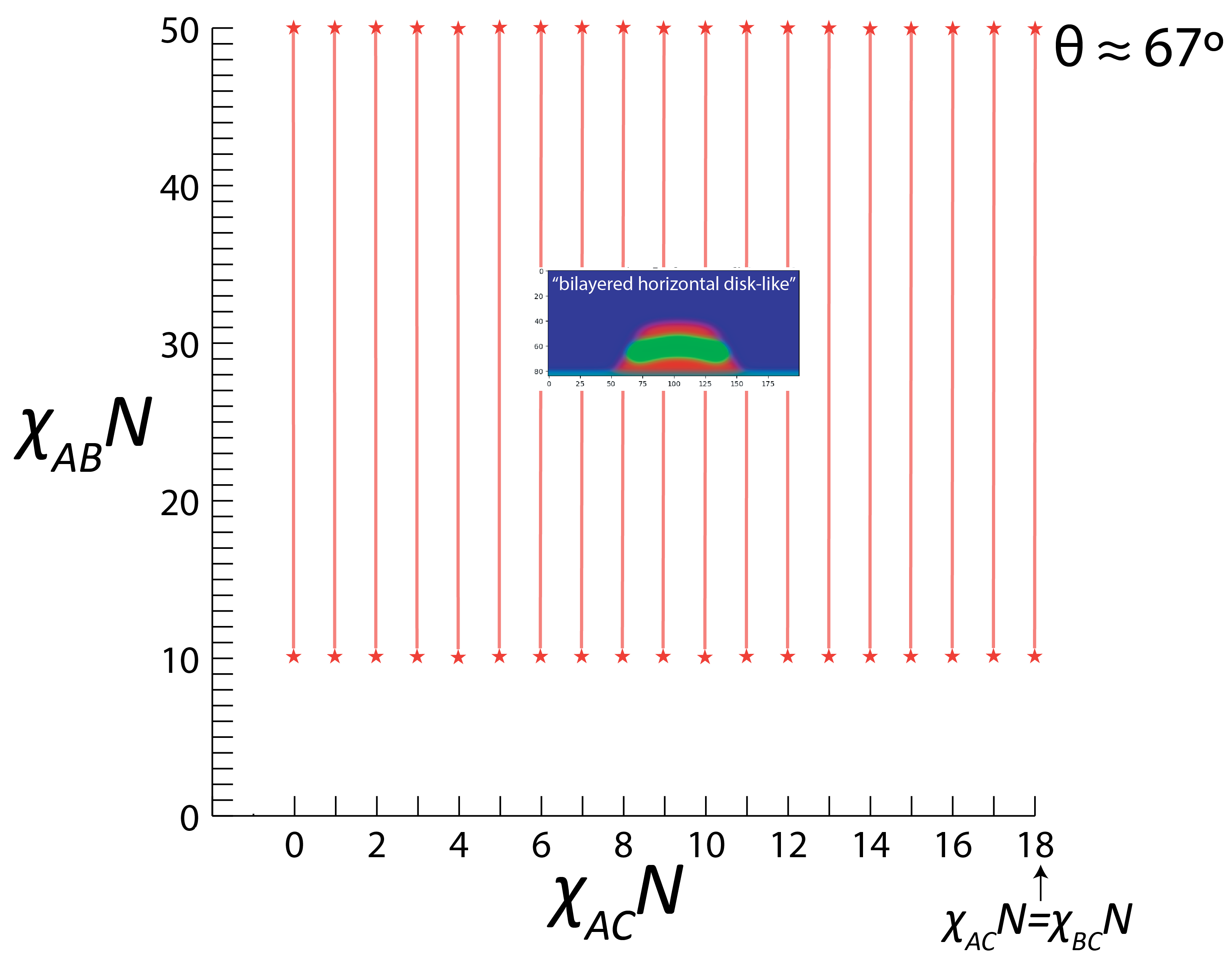}
	\caption{}
	\end{subfigure}
  \caption{Morphological diagram of nanodroplets adsorbed at a surface with "moderate" preference towards $A$ blocks ($Pref_A\approx 0.094$); the surrounding medium is either neutral or preferential to the same block ($\chi_{AC}\leq\chi_{BC}$). SCFT calculations were performed in 2D space. Diagrams (a), (b), and (c) are obtained at $\theta\approx 141\degree$, $\theta\approx 92\degree$, and $\theta\approx 67\degree$, respectively. Points of different colors represent the boundaries of different morphologies at a fixed $\chi_{AC}N$. Continuous lines between the boundary points show that this morphology is obtained at all values of $\chi_{AB}N$ between these points. For $\theta\approx141\degree$ and small $\chi_{AB}N\approx 10$, the striped ellipsoidal droplet transitioned into onion-like structure smoothly. We did not perform additional procedures to determine an exact boundary between these two morphologies in this region, since we did not carry out the iterative correction of this diagram. As a result, this transitional region is shown by points with color changing by a gradient between the color corresponding to the striped ellipsoidal droplet and the color corresponding to the onion-like structure. A characteristic snapshot for each structure as well as its short name are shown for each morphological region. $A$ blocks are shown in red, $B$ blocks are colored green, the surface-tethered $S$ homopolymers are shown in light blue, and homopolymers $C$ modeling the surrounding medium are colored dark blue.}
\label{fig_s_modpref}
\end{figure}

\begin{figure}[H]
\centering
  \begin{subfigure}{0.49\textwidth}
	\includegraphics[width=\linewidth,height=\textheight,keepaspectratio]{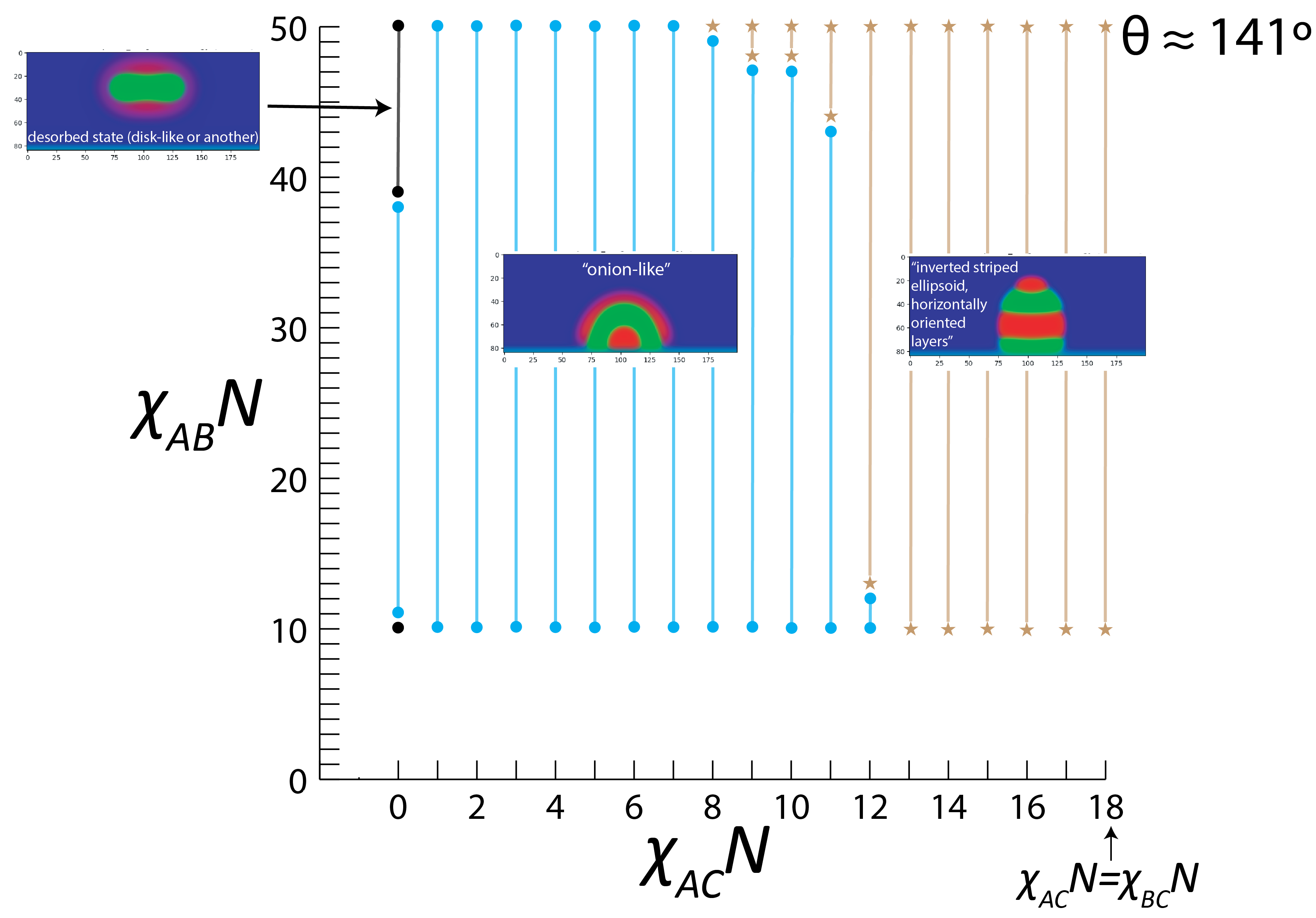}
	\caption{}
	\end{subfigure}
	\begin{subfigure}{0.49\textwidth}
	\includegraphics[width=\linewidth,height=\textheight,keepaspectratio]{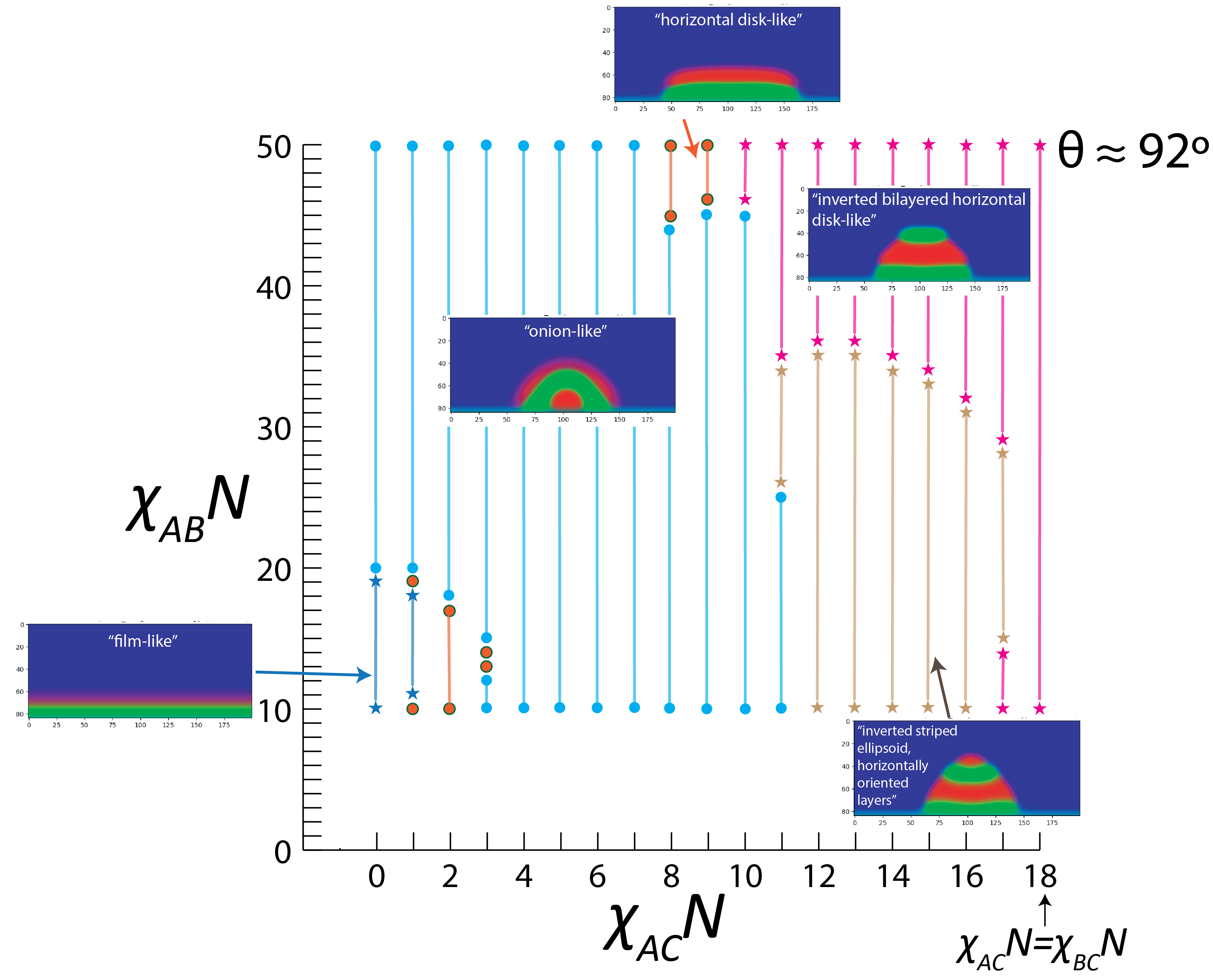}
	\caption{}
	\end{subfigure}
	\begin{subfigure}{0.49\textwidth}
	\includegraphics[width=\linewidth,height=\textheight,keepaspectratio]{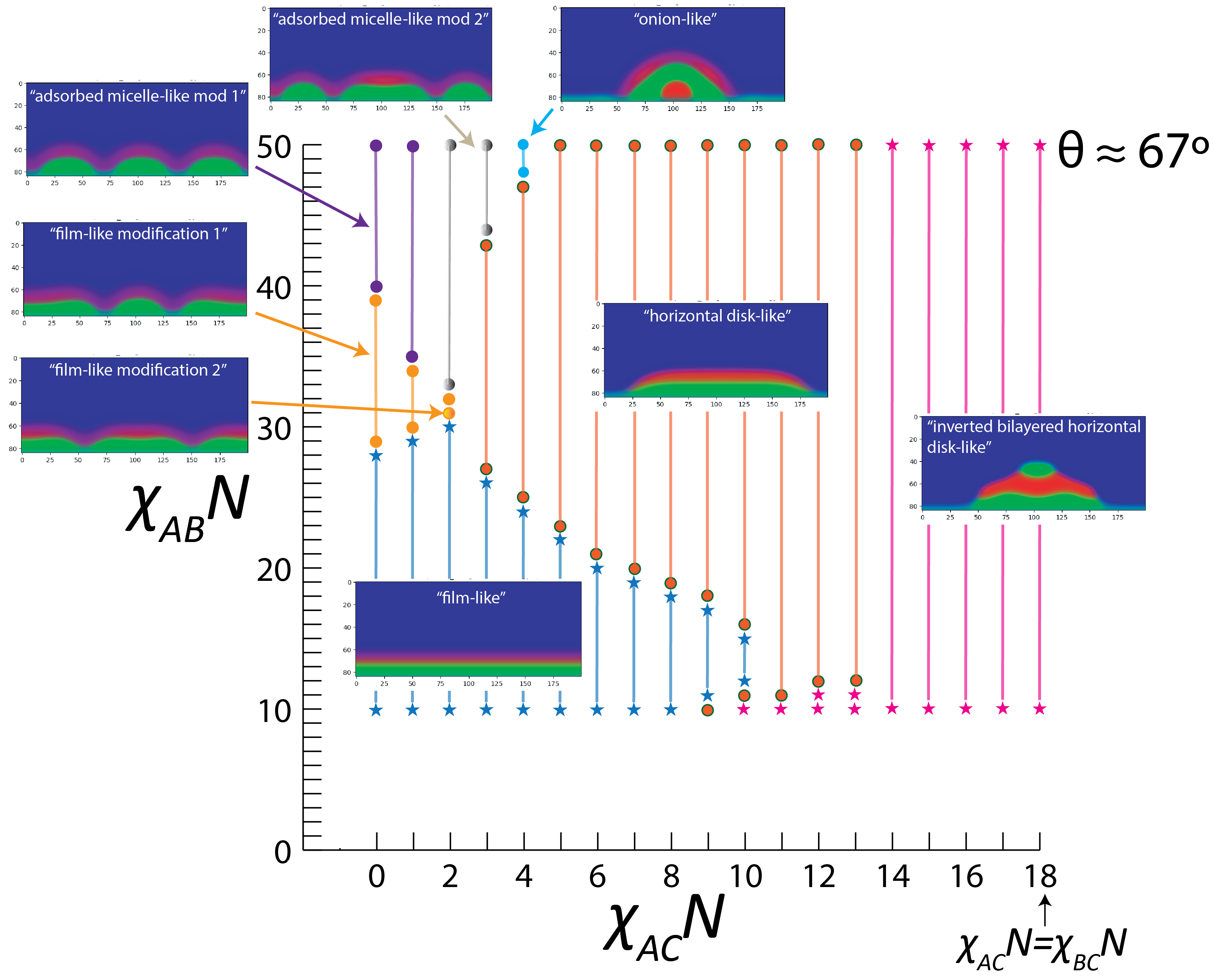}
	\caption{}
	\end{subfigure}
  \caption{Morphological diagram of nanodroplets adsorbed at a surface with preference towards $B$ blocks ($Pref_A\approx -0.064$); the surrounding medium is either neutral or preferential to the opposite block ($\chi_{AC}\leq\chi_{BC}$). SCFT calculations were performed in 2D space. Diagrams (a), (b), and (c) are obtained at $\theta\approx 141\degree$, $\theta\approx 92\degree$, and $\theta\approx 67\degree$, respectively. Points of different colors represent the boundaries of different morphologies at a fixed $\chi_{AC}N$. Continuous lines between the boundary points show that this morphology is obtained at all values of $\chi_{AB}N$ between these points. A characteristic snapshot for each structure as well as its short name are shown for each morphological region. $A$ blocks are shown in red, $B$ blocks are colored green, the surface-tethered $S$ homopolymers are shown in light blue, and homopolymers $C$ modeling the surrounding medium are colored dark blue.}
\label{fig_s_bpref}
\end{figure}


\bibliography{sample}

\end{document}